\documentclass[preprint,12pt]{elsarticle}

\usepackage{lineno}

\usepackage[T1]{fontenc}
\usepackage{graphicx}
\usepackage{xspace}      
\usepackage{amssymb}
\usepackage{amsmath}
\usepackage{amsthm}
\usepackage{mathrsfs}    
\usepackage{infer}
\usepackage[all]{xy}
\usepackage{float}      
\usepackage{array}      
\usepackage{verbatim}
\usepackage{fancyvrb}   
\usepackage{enumerate}  
\usepackage{wrapfig}    
\usepackage{stmaryrd}   
\usepackage{url}
\usepackage{xcolor}
\usepackage{tabu}
\usepackage{listings}
\usepackage{hyperref}
\usepackage[font=small,labelfont=bf]{caption}
\usepackage[commandnameprefix=always,final]{changes}

\journal{The Journal of Systems \& Software}

\begin{document}

\begin{frontmatter}



\title{Flexible and Reversible Conversion between Extensible Records and Overloading Constraints for ML}


\author[inst1]{Alvise Spanò}
\ead{alvise dot spano at unive dot it}
\affiliation[inst1]{organization={Ca' Foscari University of Venice},
            country={Italy}
            }


\begin{abstract}

Most ML-like functional languages provide records and overloading as unrelated features. Records not only represent data structures, but are also used to implement dictionary passing, whereas overloading produces type constraints that are basically dictionaries subject to compiler-driven dispatching. In this paper we explore how records and overloading constraints can be converted one into the other, allowing the programmer to switch between the two at a very reasonable cost in terms of syntactic overhead. To achieve this we introduce two language constructs, namely inject and eject, performing a type-driven syntactic transformation. The former literally injects constraints into the type and produces a function adding an extra record argument. The latter does the opposite, ejecting a record argument from a function and turning fields into type constraints. The conversion is reversible and can be restricted to a subset of symbols, granting additional control to the programmer. Although what we call inject has already been proposed in literature, making it a language operator and coupling it with its reverse counterpart represent a novel design. The goal is to allow the programmer to switch from a dictionary-passing style to compiler-assisted constraint resolution, and vice versa, enabling reuse between libraries that otherwise would not interoperate.
\end{abstract}




\end{frontmatter}

\theoremstyle{definition}
\newtheorem{notation}{Notation}[section]
\theoremstyle{plain}
\newtheorem{definition}{Definition}[subsection]
\newtheorem{theorem}{Theorem}[subsection]
\newtheorem{lemma}[theorem]{Lemma}
\newtheorem{proposition}[theorem]{Proposition}
\newtheorem{corollary}[theorem]{Corollary}


%


\newenvironment{bnf}[2][tbh]
{
\begin{table*}[#1]
    \caption{#2.}
    \centering
    \[
    \begin{array}{lcll}
}
{
    \end{array}
    \]
\end{table*}
}

\newcommand{\nontermdescr}[1]{\hspace{-8pt}$\textbf{#1}$}
\newcommand{\termdescr}[1]{\textnormal{$\textsf{#1}$}}

\newcommand{\src}[1]{\textnormal{\texttt{#1}}}

\newcommand{\olds}[1]{\oldstylenums{#1}}
\newcommand{\oldsb}[1]{{\bfseries\olds{#1}}}
\newcounter{ncomm}
\newcommand{\mnote}[1]{\stepcounter{ncomm}%
\vbox to0pt{\vss\llap{\tiny\oldsb{\arabic{ncomm}}}\vskip6pt}%
\marginpar{\tiny\bf\raggedright
{\oldsb{\arabic{ncomm}}}.\hskip0.5em {#1}}}
\newcommand{\signmnote}[3]{\mnote{\tiny\emph{#3}: \textcolor{#1}{#2}}}
\newcommand{\alvinote}[1]{\signmnote{red}{#1}{Alvi}}

\newcommand{\many}[1]{\overline{#1}}
\newcommand{\manyone}[1]{\many{#1}^{+}}

%

\newcommand{\myinferrule}[4][]{\mprset{flushleft} \inferrule*[Lab=\textsc{#2},#1]{#3}{#4}}

\newenvironment{rules}[2][tp]
{
\begin{table*}[#1]
\caption{#2}
\begin{mathpar}
}
{
\end{mathpar}
\end{table*}
}


\newcommand{\rulename}[1]{\textsc{(#1)}}

\newcommand{\ruleannot}[1]{\textnormal{\texttt{#1}}}

%


%


\newcommand{\kindannot}[2][\star]{{#2}^{#1}}

\newcommand{\rowkind}{\src{row}\xspace}
\newcommand{\rowempty}{\llparenthesis\ \rrparenthesis}
\newcommand{\rowext}[1][l]{\src{ext}_{#1}}
\newcommand{\ejectop}{\src{eject}\xspace}
\newcommand{\injectop}{\src{inject}\xspace}
\newcommand{\tejectop}{\mathcal{E}\xspace}
\newcommand{\tinjectop}{\mathcal{I}\xspace}
\newcommand{\substapp}[2][\theta]{{#1}({#2})}
\newcommand{\fvop}{\mathtt{ftv}}
\newcommand{\fvapp}[1]{\fvop(#1)}
\newcommand{\varover}[2][k]{{#2} ^ {#1}}
\newcommand{\ruleout}[2]{#1 . ~ #2}
\newcommand{\transto}{\leadsto}
\newcommand{\genop}{\simeq}
\newcommand{\isinstanceofop}{\preceq}
\newcommand{\mguop}{\mathtt{mgu}}
\newcommand{\distop}{\mathtt{dist}}
\newcommand{\distapp}[2]{\distop(#1, #2)}
\newcommand{\rankop}{\mathtt{R}}
\newcommand{\rankapp}[1]{\rankop(#1)}
\newcommand{\idset}{\mathbb{X}}

\newcommand{\compactop}{\mathbf{C}}
\newcommand{\compactapp}[2]{\compactop({#1}; {#2})}
\newcommand{\csolveop}{\mathbf{S}}
\newcommand{\csolveopr}{\mathbf{\mu}}

\newcommand{\powerset}[1]{\mathscr{P}(#1)}
\newcommand{\fresh}[1]{\mathbf{#1}}
\newcommand{\convop}{\approx}
\newcommand{\convr}[1]{\mathcal{R}^{-1}({#1})}
\newcommand{\convc}[1]{\mathcal{R}({#1})}
\newcommand{\wdash}{\vdash_{w}}
\newcommand{\infout}[4]{#1. ~ #2 \transto #3 \rhd  #4 }
\newcommand{\instantiate}[1]{\mathtt{inst}(#1)}
\newcommand{\hashop}{\mathcal{H}}
\newcommand{\hashapp}[1]{\hashop(#1)}

%

\definecolor{numbercol}{rgb}{0.8,0.2,0.2}
\definecolor{keywordcol}{rgb}{0.8,0.0,0.2}
\definecolor{stringcol}{rgb}{0.5,0.2,0.4}
\definecolor{commentcol}{rgb}{0.0,0.4,0.1}
\definecolor{codepurple}{rgb}{0.58,0,0.82}
\definecolor{verylightgray}{rgb}{0.98,0.98,0.98}

\lstdefinestyle{mystyle}{
    backgroundcolor=\color{verylightgray},
    commentstyle=\color{commentcol},
    keywordstyle=\color{keywordcol},
    keywordstyle=[2]{\color{codepurple}},
    numberstyle=\color{numbercol},
    stringstyle=\color{stringcol},
    basicstyle=
        \ttfamily\small
        \lst@ifdisplaystyle\footnotesize\fi,        
    breakatwhitespace=false,         
    breaklines=false,                 
    captionpos=b,                    
    keepspaces=true,                 
    numbersep=5pt,                  
    showspaces=false,                
    showstringspaces=false,
    showtabs=false,                  
    tabsize=2
}

\lstset{style=mystyle}

\lstnewenvironment{lwcode}
{\lstset{language=caml,
    keywordstyle=[2]{\color{codepurple}},
    morekeywords={over,overload,eject,inject,?,val,where},
    morekeywords=[2]{foldr,foldl,map},
    morecomment=[s]{//}{\^^M},
    }}
{}

\lstnewenvironment{haskellcode}
{\lstset{language=haskell,
    deletekeywords={sum,zero,map,foldl,foldr},
    morekeywords={inject,eject,from},
    morekeywords=[2]{foldr,foldl,map},
}}
{}

\nolinenumbers

\section{Introduction}
\label{sec:intro}

In this paper we introduce a language design aimed at integrating two programming practices that are typically considered orthogonal: explicit dictionary passing and compiler-driven resolution of overloading constraints.
Dictionary passing in functional languages is typically achieved through records.
Records come in multiple flavours: ML-like languages historically featured heavyweight records \cite{cardelli1985understanding}, i.e., product data types that must be explicitly defined by the programmer, consisting of global label names and supporting neither extensibility nor polymorphism.
In the literature more powerful record systems have been proposed to overcome such limitations: lightweight records require no annotations or declarations by the programmer, the shape of records is inferred from the program text and labels can be reused across multiple record types.
Extensible records enhance this even further by adding extensibility and polymorphism, either via subtyping \cite{Cardelli1990,cardelli1992extensible,jategaonkar1993type} or row types \cite{Wand1987CompleteTI,wand1991type,Remy:1989:TCR:75277.75284}.
Language extensions implementing extensible records exist for Haskell, OCaML and other MLs \cite{jones1999lightweight,gaster1996polymorphic,gaster1998records}.

Overloading, on the other hand, brings ad-hoc polymorphism to the table \cite{kaes1988parametric,Wadler:1989:MAP:75277.75283}.
The most powerful form of open-world context-dependent overloading is traditionally based on type constraints, i.e. qualified types \cite{jones1992theory,Jones:1995:SIQ:224164.224198} representing dictionaries of functions that are statically resolved and dispatched by the compiler.
Haskell and Scala provide notorious implementations where overloadable symbols are grouped into global named collections called type classes \cite{hall1996type,Jones97typeclasses} and traits \cite{odersky2004overview,Odersky2014} respectively.
Alternative overloading systems exist in the literature, such as System CT \cite{Camarao:1999:TIO:646189.683411,Camarao:2004:CSO:1013963.1013974} and System O \cite{Odersky:1995:SLO:224164.224195}, where overloadable symbols are not grouped into type classes, but are rather stand-alone names that can be individually overloaded by the programmer.
In such cases, type constraints appear as a flat series of overloadable identifiers along with their types.
For our proposal, we favour this approach for a number of reasons that will be motivated later in this paper.
We refer to this form of type constraint as \emph{fine-grained}.

Extensible records and overloading rely on two distinct dispatching mechanisms: dynamic dispatching is required to dispatch record fields at runtime \cite{bruce1999comparing}, while static dispatching of overload instances takes place at compile time \cite{Wadler:1989:MAP:75277.75283,Jones97typeclasses}.
The two mechanisms are orthogonal, support different forms of polymorphism and employ rather different coding styles.
Such differences discourage reuse across the two, due to the significant amount of code required for interoperation.

In this paper we propose a language design that enhances code reuse between these two styles, making records and overloading interoperate easily from a programmer's perspective.
We achieve this by introducing a pair of language constructs that transform a function picking a record argument into a constrained function that can be solved by the compiler, and the other way around.

\subsection{A problem of interoperation}

\begin{figure}
\centering
\begin{haskellcode}
-- Library 1: ring-like objects, uses records as dictionaries.
data Ring a = { zero :: a;
                one  :: a; 
                add  :: a -> a -> a;
                mult :: a -> a -> a }

-- provides only a way for multiplying elements in a list
multR :: Ring a -> [a] -> a

-- Library 2: semigroup supporting addition, uses type classes.
class Semigroup a where
    zero :: a
    add :: a -> a -> a

-- provides only a way for summing elements in a list.
sum :: Semigroup a => [a] -> a 

-- The following is pseudo-Haskell code, as the language
-- does not fully support two required features: local 
-- resolution and local instances.

-- We want to mix the two libraries, making them interoperate
-- to perform multiplications and sums over a list of lists
-- of elements. We also want to perform computations using
-- a list of ring dictionaries. For each ring, we want to
-- multiply the elements in each inner list using Library 1,
-- then sum all the results using Library 2.
msums :: [Ring a] -> [[a]] -> [a]
msums rings lists = 
    let f ring =
        -- Local instances would be required for this
        let instance Semigroup a where 
            -- stub record fields to instance names
            zero = ring.zero
            add = ring.add
        -- Local resolution would be needed now,
        -- preventing the constraint from being lifted
        in sum (map (multR ring) lists)
    in map f rings                           
\end{haskellcode} 
\caption{Mixing manual dictionary-passing with constraint resolution can be problematic in Haskell.
The snippet is hypothetical, as it shows a workaround that would require Haskell to support local instances and local resolution.
Notably, even if such language extensions were fully supported, a considerable amount of wrapper code would still be required.}
\label{fig:problem-haskell}
\end{figure}

Figure \ref{fig:problem-haskell} shows pseudo-Haskell code attempting to integrate two libraries using different styles.
The code relies on language extensions that are not fully supported by the Haskell language and would be required for interoperation.
The first library is based on records representing ring-like structures.
It provides a function \src{multR} that multiplies all elements in a list using the operators defined in the \src{Ring} record passed as argument.
The second library is based on type classes representing additive semigroups.
It provides a function \src{sum} that sums all elements of a list using the overloadable symbols in the type class.

Function \src{msums} then attempts the integration of the two libraries.
It wants to multiply and then sum all elements in a list of lists using a different \src{Ring} each time.
Library 1 can only multiply and Library 2 can only sum: co-operation is necessary across two different styles.
The problem is Library 1 uses a record-passing style, whereas Library 2 relies on the compile-time resolution of type constraints.
In Haskell the code in Figure \ref{fig:problem-haskell} cannot be written as it requires local instances and local resolution of the type constraint produced by each \src{sum} call.
Without such extensions, the type constraint is propagated upwards, and that leads to the constrained type $\src{msums} :: \src{Semigroup} ~ a \Rightarrow [\src{Ring} ~ a] \rightarrow [[a]] \rightarrow [a]$, meaning only one instance of \src{Semigroup} is dispatched statically for the whole function, rather than a different one for each inner call to \src{sum} as originally intended.

It is interesting to point out, however, that even with the appropriate Haskell extensions, the problem of integrating records and type classes would remain, due to the amount of wrapper code required to make the two styles cooperate, as the snippet shows.
Figure \ref{fig:problem-haskell-jects} presents the same scenario of Figure \ref{fig:problem-haskell} but it introduces our proposal for making the two libraries interoperate, two language operators named \injectop and \ejectop. The former converts a constrained function into a function picking an explicit record argument, and the latter does the opposite. As it will turn out, in Haskell these conversions cause several complications that prevent \injectop and \ejectop from being effectively added to the language, hence the need to design a custom language that is tailored to them.

\subsection{Our Proposal}
\label{sec:solution}

We introduce a pair of unary operators, \injectop and \ejectop, that are second-class citizens of the expression syntax.
Intuitively, given a constrained value $f : \pi \Rightarrow \tau$, where $\pi = \{~ x_1 : \tau_1; ~ .. ~; x_n : \tau_n ~\}$ is a constraint set, then expression $\injectop ~ f : \pi \rightarrow \tau$ transforms $f$ into an unconstrained function where $\pi$ is a new record argument that is isomorphic to the constraint set.
This allows the programmer to perform explicit dictionary passing.
We call this \emph{injection}, because it injects the constraint $\pi$ \emph{into} the type by adding a new arrow.
Existing proposals in the literature call this operation instance arguments \cite{Devriese:2011:BST:2034773.2034796} and explicit dictionary application \cite{winant2018coherent}, as it somewhat reveals to the programmer the underlying dictionary passing mechanism that was first introduced by \cite{kaes1988parametric,Wadler:1989:MAP:75277.75283} and implemented in Haskell to pass instances to constrained functions.

We call \emph{ejection} the reverse operation, as it ejects a record argument \emph{out} of an arrow type by consuming the domain and adding a new constraint set.
Let $g$ be the unconstrained function $\injectop ~ f : \pi \rightarrow \tau$, with $\pi = \{~ x_1 : \tau_1; ~ .. ~; x_n : \tau_n ~\}$ being a record type, then expression $\ejectop ~ g : \pi \Rightarrow \tau$ yields back to $f$ by erasing the record argument and turning all fields into a new constraint set that is isomorphic to the record argument.

\begin{figure}
\centering
\begin{haskellcode}
-- Library 1: ring-like objects, uses records as dictionaries
data Ring a = { zero :: a;
                one  :: a; 
                add  :: a -> a -> a;
                mult :: a -> a -> a }

multR :: Ring a -> [a] -> a

-- Library 2: semigroup supporting addition, uses type classes
class Semigroup a where
    zero :: a
    add :: a -> a -> a

sum :: Semigroup a => [a] -> a 

-- Now we make the two libraries interoperate

-- we first produce a new version of the multR 
-- without the record argument
mult :: Ring a => [a] -> a -- arrow (->) became a (=>)
mult = eject multR

-- then a version of sum function picking
-- an extra record argument
sumR :: Semigroup a -> [a] -> a -- arrow (=>) became a (->)
sumR = inject sum

-- finally we use the two new stubs
msums :: [Ring a] -> [[a]] -> [a]
msums rings lists = 
    let f = sumR (map mult lists)
    in map f rings                     
\end{haskellcode} 
\caption{The snippet shows how to make two libraries based on different programming practices interoperate through a combined use of injection and ejection. Despite some technical complications that arise in Haskell, mainly due to heavyweight records and monolithic type classes, the code gives an intuition of what the \injectop and \ejectop operators are capable of.}
\label{fig:problem-haskell-jects}
\end{figure}

Additionally, injection and ejection support restriction over a subset of symbols, for example $\injectop ~ x_1 ~ \src{in} ~ f : \{~ x_2 : \tau_2; ~ .. ~; x_n : \tau_n ~\} \Rightarrow \{ ~ x_1 : \tau_1 ~\} \rightarrow \tau$ and $\ejectop ~ x_1 ~ \src{in} ~ g : \{ ~ x_1 : \tau_1 ~\} \Rightarrow \{~ x_2 : \tau_2; ~ .. ~; x_n : \tau_n ~\} \rightarrow \tau$.
This allows fine-grained manipulation of record fields and constraints.

Syntactically, \injectop and \ejectop are first-class language operators, thus supporting any form of expression as operands and are not limited to operating on function names.
Inject works on expressions of any type as long as the constraint set is non-empty in the current context so that it can be turned into a new record argument. Symmetrically, eject requires the type of its expression operand to match an arrow type with a record on the domain.
This allows for mixed uses of the operators at any nesting level within expressions, e.g. \src{(fun x -> inject b in (eject a b in (fun r -> r.a + r.b + x))) 3 \{ b = 4 \}} has type $\{~ a : \src{int} ~\} \Rightarrow \src{int}$.
Proposals in the literature including those mentioned above \cite{Devriese:2011:BST:2034773.2034796,winant2018coherent} do not support nesting at the syntactic level.
Additionally, what we call ejection represents a novelty on its own, introducing a reversible mechanism.
To the best of our knowledge, this is unprecedented in the literature, as it introduces a novel reversible constraint-to-record conversion system that can be a valuable programming tool, enabling new forms of code reuse by making the world of records and the world of overloading communicate.

\subsection{Contribution}

Understanding the contribution of this paper requires an explicit mention of where it collocates in such a crowded design space.
Our system relies on two underlying subsystems.
The overloading subsystem is a hybrid inspired by other overloading systems \cite{Camarao:1999:TIO:646189.683411,Jones97typeclasses}, and integrates features that are known in the literature with features that are novel to some extent, such as local instances and local scoping of overloaded symbols. This is not the focus of our contribution, though, and will be discussed in Section \ref{sec:overview-overloading}.

The main contribution of this paper is the introduction of a flexible way for converting records into overloading constraints (and the other way around) through a pair of first-class language operators, the already mentioned \ejectop and \injectop.
Being novel language features, this preliminary work is aimed at presenting a first formulation of the system as an extension to the ML functional language. Section \ref{sec:ject} explores some of its potential applications by examples, which include (but are not limited to) enhancing the interoperability between libraries or pre-existing code written in different styles.
In Section \ref{sec:type-system} we provide a formalization of our type system that wants to be as self-contained as possible, covering all aspects of the proposed design, including the details of the two underlying subsystems, namely records and overloading.
Injection and ejection are built on top of those and cannot live as stand-alone features, since their purpose is to make the two subsystems interoperate. The result is a layered and thick design for which we provide a full type inference algorithm in Section \ref{sec:type-inference}, as a guidance for potential implementations willing to introduce our operators in their languages.

As a preliminary and introductory work, the paper presents a first effort towards a full formalization of the system, sketching the most relevant proofs in Section \ref{sec:soundness}. A more comprehensive theoretical framework including the proofs of soundness and coherence of the whole system will be the target of a future publication further exploring the injection and ejection and their implications.

\subsection{Originality}

With such a layered proposal, the originality of this work has to be carefully outlined.
Converting type constraints into explicit record passing (which is what we call injection) has already been invented and studied for Haskell \cite{winant2018coherent} and Adga \cite{Devriese:2011:BST:2034773.2034796}, though with a rather different approach. Our system reformulates that as a first-class language operator for an extension of the ML functional language.
A more in-depth comparison is provided in the Related Works in Section \ref{sec:related}.

Moreover, injection is only half of the picture.
Its reverse counterpart, ejection, is something novel in the literature and in programming languages in general.
The combined use of the \injectop and \ejectop operators at any nesting level in the code provides the programmer with a reversible conversion mechanism that enables new forms of code reuse that have yet to be explored by the programming languages community, as discussed in Section \ref{sec:ject}.
    
In addition, injection and ejection support \emph{restriction} over a set of identifiers. 
This allows the programmer to specify which overloaded identifiers or record fields shall be manipulated by \injectop or \ejectop.
This is thoroughly novel and enhances code reuse even further.
Section \ref{sec:restricted-ject} discusses such aspects in detail.

\section{Preliminaries}
\label{sec:preliminaries}

The design proposed in this paper is an ML-like language extended with a form of open-world context-dependent overloading based on fine-grained constraints and extensible records, on top of which we add \injectop and \ejectop.
The reason behind this choice is mostly due to complications arising from heavyweight record types and monolithic type classes à la Haskell.
The problem is that \injectop and \ejectop must generate new record types and new type classes, respectively.
Name clashes would therefore arise among record fields and global overloaded names, possibly producing unwanted shadowing and introducing the need to re-bind record fields to type class members.

Take into consideration the following example: a simple injection takes place, converting the \src{Semigroup} constraint into a homonymous record type. Such record type must be already defined, though, or it must be generated automatically.

\begin{haskellcode}
class Semigroup a where
  unit : a
  plus : a -> a -> a

sum :: Semigroup a => [a] -> a
sum [] = unit
sum (x : xs) = plus x (sum xs)

-- cannot inject! 
-- Semigroup is a type class, not a record type
rsum :: Semigroup a -> [a] -> a
rsum = inject sum
\end{haskellcode}

Generating a record type mimicking the type class might introduce unwanted name clashes between label names.

\begin{haskellcode}
type AnotherRecord = { add :: a -> a -> a; sub : a -> a -> a }

-- this record type must be generated by inject
-- though, label 'add' is already defined in the record above
type Semigroup a = { zero :: a; add :: a -> a -> a }

-- now injection is possible
-- but a name clash has occurred
rsum :: Semigroup a -> [a] -> a
rsum = inject sum
\end{haskellcode}

Even though injection is possible after generating the corresponding record type, label \src{add} clashes with a pre-existing record, meaning a lot of care must be put when dealing with injection: record labels require fully qualified access.
Analogously, ejection would require the generation of type classes mimicking record types, hence the risk of name clashes between global overloadable symbols.
To avoid these complications we preferred a different approach: extensible records allowing the reuse of field names and fine-grained overloading allowing shadowing in case of duplicate definitions.
Additionally, support for local instances and local resolution is required in order to make \injectop and \ejectop work as intended at any nesting level within expressions.
The introductory example in Figure \ref{fig:problem-haskell} already showed the need for such features.

\subsection{Overview of extensible records}

Among the many systems for extensible records available in literature \cite{gaster1996polymorphic,Remy:1989:TCR:75277.75284,cardelli1992extensible} we adopt extensible records with scoped labels \cite{extensible-records-with-scoped-labels} because of its balance between simplicity and expressivity.
It adds little complexity to the classic Hindley-Milner type inference algorithm by only affecting unification and introduces a basic kind system.

A sample function using this system for summing all elements of a list using dictionary passing is:
\begin{lwcode}
let rec sum r l =
  match l with
  | [] -> r.zero
  | x :: xs -> r.add x (sum r xs)
\end{lwcode}

The inferred type $\src{sum} : \{~ \src{zero} : \alpha; ~ \src{add} : \beta \rightarrow \alpha \rightarrow \alpha ~|~ \kindannot[\rowkind]{\gamma} ~\} \rightarrow \beta ~ \src{list} \rightarrow \alpha$ includes a row type $\kindannot[\rowkind]{\gamma}$ that represents the unknown tail of the record as a type variable $\gamma$ of kind \src{row}.
This is standard practice in row type systems \cite{Wand1987CompleteTI} and enables passing to the \src{sum} function a record argument consisting of the two fields \src{zero} and \src{add}, plus an arbitrary amount of extra fields. The row type variable $\gamma$ unifies with such extra fields, factually simulating subtyping without the need to introduce an actual subtype relation among types.



\subsection{Overview of the overloading subsystem}
\label{sec:overview-overloading}

As anticipated, our take on overloading is not the focus of this paper, although it is a necessary ingredient for exposing the main contribution, i.e., injection and ejection. It integrates existing features in the literature in a way that is novel to some extent, leaving the need to provide a detailed overview.
We adopt a fine-grained open-world context-dependent overloading system requiring type signatures for overloaded symbols like in \cite{Camarao:1999:TIO:646189.683411} and \cite{Odersky:1995:SLO:224164.224195}.
This is equivalent to granular type classes consisting of a single overloadable symbol, thus principal typability is preserved in the same way as in any parametric overloading system \cite{kaes1988parametric}.

\begin{lwcode}
overload add : 'a -> 'a -> 'a

let twice x = add x x // add can be used even without instances
\end{lwcode}

Although there is no instance for \src{add} when \src{twice} is defined, the inferred type is $\{~ \src{add} : \alpha \rightarrow \alpha \rightarrow \alpha ~\} \Rightarrow \alpha \rightarrow \alpha$.

Instances are introduced by means of a special letover-binding\footnote{Assume that the binary operator $\src{(+)} : \src{int} \rightarrow \src{int} \rightarrow \src{int}$ is the monomorphic addition for integers.}:

\begin{lwcode}
let twice_one = twice 1 // unsolved as no instance is available
let over add a b = a + b // instance of add for integers
let four = twice twice_one // solved, evaluates to 4
\end{lwcode}

The inferred types show that \src{twice\_one} is kept unsolved and exhibits a monomorphic constraint, while \src{four} is solved and evaluates to the ground value 4 because an instance for \src{add} exists at that point in the program.
Unsolved monomorphic constraints are not rejected and produce constrained monomorphic values.
$$
\begin{array}{rcl}
\src{twice\_one} & : & \{~ \src{add} : \src{int} \rightarrow \src{int} \rightarrow \src{int} ~\} \Rightarrow \src{int}
\\
\src{add} & : & \src{int} \rightarrow \src{int} \rightarrow \src{int}
\\
\src{four} & : & \src{int}
\end{array}
$$

We call \emph{resolution} what \cite{Jones97typeclasses} defines as \emph{context reduction}, though we refer to single constraints, not whole type classes.
A type constraint $o_1 : \tau_1$ for an overloaded symbol $o_0 : \tau_0$, where $\tau_1 = \substapp{\tau_0}$ for some substitution $\theta$, gets \emph{solved} when a \emph{best-fitting} instance is available among those available in the environment.
That is, some instance $o : \tau$ such that the \emph{type distance} between $\tau$ and the constraint type $\tau_0$ is minimal compared to other instances.
We are giving a formal description of this mechanism later in Section \ref{sec:type-system}.
When multiple instances are a best fit, i.e., when there is not just one exhibiting a minimum type distance, constraints are kept unsolved.
An implementation may freely introduce custom strategies for dealing with unsolvability, albeit that is beyond the scope of this paper.

Instances can have type constraints too:
\begin{lwcode}
overload (=) : 'a -> 'a -> 'a
overload (<) : 'a -> 'a -> 'a
overload (<=) : 'a -> 'a -> 'a
let over (<=) x y = x < y || x = y  // constrained instance
\end{lwcode}

Assuming the logical \emph{or} operator $\src{(||)} : \src{bool} \rightarrow \src{bool} \rightarrow \src{bool}$ exists, the instance for \src{(<=)} assumes the constrained polymorphic type $\{~ \src{(=)} : \alpha \rightarrow \alpha \rightarrow \alpha;~ \src{(<)} : \alpha \rightarrow \alpha \rightarrow \alpha ~\} \Rightarrow \alpha \rightarrow \alpha \rightarrow \alpha$.
Generalization for letover-bindings works exactly as for ordinary let-bindings, producing constraints in a natural way.
Overload declarations consist of unconstrained types, though, as in most overloading systems including Haskell type classes.
In general, an instance $o_1 : \pi_1 \Rightarrow \tau_1$ for an overloaded symbol declared as $o_0 : \tau_0$ is valid when a substitution $\theta$ exists such that $\tau_1 = \substapp{\tau_0}$ without taking into account the instance constraint set $\pi_1$.

Local instances are allowed as in \cite{Camarao:1999:TIO:646189.683411}.
Unsolved constraints in conjunction with local instances lead to a form of dynamic scoping:
\begin{lwcode}
let one =
    let over add a b = a * b // local instance
    in twice twice_one // solved locally, evaluates to 1
\end{lwcode}

The local instance solves the monomorphic constraint of \src{twice\_one}, computing $1 * 1 = 1$, as well as the constraint of \src{twice}, overall leading to $1$ again.
Intuitively, a type constraint gets solved only if a best-fitting instance is available without ambiguities, otherwise it is simply kept unsolved, rather than producing an error.

\subsection{Implicit parameters and interactions with overloading}
\label{sec:interactions-implicits}

Implicits parameters, or simply \emph{implicits}, are function parameters that do not appear as explicit parameters in the function signature, and are applied automatically by the compiler through a type-driven resolution algorithm. Overloading and implicits rely on the same resolution mechanism and are therefore subject to the same dictionary translation \cite{Lewis:2000:IPD:325694.325708}.


The reason why implicits are needed by our system is that they serve as a fallback mechanism for ejection when a record field name does not correspond to an overloadable symbol in the scope, as discussed in Section \ref{sec:ejection}.
Our system supports implicits by prefixing a question mark to a variable identifier.
Implicits require no principal type declaration, similarly to the Haskell extension for implicit parameters based on \cite{Lewis:2000:IPD:325694.325708}, and produce constraints that can be freely mixed with overloading constraints.

\begin{lwcode}
let twice x = ?add x x  // use of implicit

let nine =
    let add = ( * ) // simple binding solves implicit
    in twice 3      // solved locally

overload add : 'a -> 'a -> 'a // now add becomes overloaded

let over add = (+)  // instance for add
let six = twice 3   // instances can solve implicits too
\end{lwcode}

The type of \src{twice} is $\{~ \src{?add} : \alpha \rightarrow \alpha \rightarrow \beta ~\} \Rightarrow \alpha \rightarrow \beta$ and is not exactly equivalent to the type of \src{twice} in Section \ref{sec:overview-overloading}, being it more general.
However, the point is that question-marked constraints can be solved by plain let-bindings as well as overload instances.
This is true even if the overload declaration occurs below, as in the resolution of \src{six}.

An implicit \src{?o} occurring in a scope where the name \src{o} is overloaded is treated independently of the principal type of \src{o}, producing a separate question-marked constraint with a fresh type variable.
The following shows mixed use of implicits and overloads\footnote{Assume that the binary operator \src{\^} is the string concatenation operator.}:

\begin{lwcode}
overload pretty : 'a -> string

let rec pretty4 l =
  match l with
  | [] -> ""
  | (x, y, z, w) :: ss ->
     pretty x ^ ?pretty y ^ pretty z
     ^ ?pretty w ^ pretty4 ss
\end{lwcode}

Type $\src{pretty4} : \{~  \src{pretty} : \alpha \rightarrow \src{string}; ~ \src{?pretty} : \beta \rightarrow \src{string}; ~ \src{pretty} : \gamma \rightarrow \src{string}; ~\} \Rightarrow (\alpha * \beta * \gamma * \beta) ~ \src{list} \rightarrow \src{string}$ reveals that the type of the implicit \src{?pretty} was inferred separately.
Moreover, multiple occurrences of the same implicit \src{?pretty} do not produce multiple constraints, whereas multiple occurrences of the same overloaded identifier \src{pretty} do.
Reusing an implicit parameter has the same effect as reusing a lambda parameter.
This is a deliberate design choice as it is less error-prone for the programmer, who may accidentally create undesired multiple implicit parameters by reusing the same name.
These behaviours are discussed in detail in Section \ref{sec:type-system}.

\section{Injection and Ejection in Detail}
\label{sec:ject}

As a foreword, it is interesting to point out that injection and ejection are not just a way to reveal the underlying dictionary-passing mechanism. 
Dictionary passing does not directly involve records, though, and is typically implemented via nested lambda abstractions and the consequent application of instances.
As will turn out, injection and ejection fold and unfold such nested lambdas into records.

\subsection{Injection}
\label{sec:injection}

Consider the code in Figure \ref{fig:inject-translation-example}: the right-hand side shows the translation performed by our type rules formalized in Table \ref{tab:type-rules-2} of Section \ref{sec:type-system}.
Most type systems featuring constrained or qualified types implement a compile-time code translation strategy embedded within the type rules \cite{Jones:1995:SIQ:224164.224198} for transforming the input program into a simpler program while preserving types and other soundness properties (details in the Section \ref{sec:soundness}).
Principal type declarations of overloadable symbols are erased at compile type and do not appear in the translated code.
Function \src{sum} in Figure \ref{fig:inject-translation-example} gains two extra parameters when translated, one for each overloaded symbol.
Even though constraints are fine-grained, as opposed to the more common approach of type classes, our system implements the same dictionary-passing translation mechanism invented by \cite{kaes1988parametric,Wadler:1989:MAP:75277.75283}.

\begin{figure}
\centering
\begin{tabular}{p{16em}|p{15em}}
\textbf{Inject Example}
&
\textbf{Translation}
\\
\hline
\begin{lwcode}
overload zero : 'a
overload add : 'a -> 'a -> 'a

let rec sum l =
  match l with
  | [] -> zero
  | x :: xs ->
      add x (sum xs)
    
let concat_strings =
    (inject sum)
        { zero = "";
          add = (^) }
\end{lwcode}
&
\begin{lwcode}



let rec sum zero add l =
  match l with
  | [] -> zero
  | x :: xs ->
    add x (sum zero add xs)
            
let concat_strings =
 (fun r -> let zero = r.zero
           let add = r.add
           in sum zero add)
   { zero = "";
     add = (^) }
\end{lwcode}
\end{tabular}
\caption{Example of injection and typing-time translation. Dictionary passing relies on extra arguments \src{zero} and \src{add} in the translated function \src{sum} on the right. Below, injection introduces a new record argument \src{r} and locally binds all overloaded names to the fields of \src{r}. This mimics the resolution of the constraints of \src{sum} by providing a local set of symbols to be passed as a dictionary.}
\label{fig:inject-translation-example}
\end{figure}

Function $\src{concat\_strings}$ has type $\src{string list} \rightarrow \src{string}$ due to the explicit dictionary passing taking place in the body.
The expression \src{inject sum} on the left side of the application is translated into a lambda abstraction over a record parameter \src{r} that contains all the extra arguments to be passed to function \src{sum}.
The basic idea behind the translation is to let-bind record fields to overloaded symbol names; dictionary passing will do the rest.
Extensible records are necessary to make this mechanism scale at any nesting level within programs.






\subsection{Ejection}
\label{sec:ejection}

A basic example of \ejectop translation is depicted in Figure \ref{fig:eject-translation-example}.
Function \src{sum} is just based on records, and translation does not change it.
Ejection erases the record argument of function \src{sum} and converts its fields into constraints, which are eventually converted by the translation into additional arguments due
to dictionary passing.

\begin{figure}[h]
\centering
\begin{tabular}{p{16em}|p{15em}}
\textbf{Eject Example}
&
\textbf{Translation}
\\
\hline
\begin{lwcode}
let rec sum r l =
    match l with
    | [] -> r.zero
    | x :: xs ->
        r.add x (sum r xs)

overload zero : 'a
overload add : 'a -> 'a -> 'a

let osum = eject sum
\end{lwcode}
&
\begin{lwcode}
let rec sum r l =
    match l with
    | [] -> r.zero
    | x :: xs ->
        r.add x (sum r xs)



            
let osum = fun zero add ->
    sum { zero = zero;
          add = add }
\end{lwcode}
\end{tabular}
\caption{Example of ejection and translation performed by type rules. Ejection restores dictionary passing by adding lambda arguments \src{zero} and \src{add}, one for each field of the record parameter \src{r} of function \src{sum}. This is then erased by passing a record value, where all fields are bound to overloaded names.}
\label{fig:eject-translation-example}
\end{figure}


The type of $\src{osum}$ is $\{ ~ \src{zero} : \alpha; ~ \src{add} : \alpha \rightarrow \alpha \rightarrow \alpha ~ \} \Rightarrow \alpha ~ \src{list} \rightarrow \alpha$.
The trick is to rebind record fields to overloaded names being passed as arguments: in such a way the translated function becomes a regular constrained function implementing dictionary passing so that the resolution system can deal with it.

One major problem with ejection is that record field names being transformed into constraints must already be defined as overloadable symbols in the scope.
When no overload declaration exists for a given field, an implicit is generated.
Consider the following nested use of eject, in which fields \src{times} and \src{k} of the lambda record parameter \src{r} are undefined symbols in the outer scope, thus are turned into implicits:

\begin{lwcode}    
let multy = map (eject fun r x -> r.times r.k x)
\end{lwcode}

The type of $\src{multy}$ is $\{ ~ \src{?k} : \alpha; ~ \src{?times} : \alpha \rightarrow \beta \rightarrow \gamma ~ \} \Rightarrow \beta ~ \src{list}  \rightarrow \gamma ~ \src{list}$.
Ejection literally \emph{ejects} a record argument, lifting all its fields to the constraint set, which is eventually solved statically by the compiler.

\subsection{Combining injection and ejection}
\label{sec:combining-jects}

In Figure \ref{fig:ejection-big} we present a solution to the problem originally introduced in Figure \ref{fig:problem-haskell} using our language design. The combined use of injection and ejection has already been presented for Haskell in Figure \ref{fig:problem-haskell-jects}, but now problems with name clashing do not arise thanks to extensible records and fine-grained overloading.

\begin{figure}[h]
\centering
\begin{lwcode}
overload zero : 'a
overload plus  : 'a -> 'a -> 'a

let rec sum l =
    match l with
    | [] -> zero
    | x :: xs -> plus x (sum xs)

let multR ring = foldl ring.times ring.one

let msums rings list =
    let f ring =
         // instances for sum contraints
        let over zero = ring.zero
        let over plus = ring.plus    
        // bindings for implicits introduced by eject
        let times = ring.times 
        let one = ring.one
        in sum (map (eject multR) list)
    in map f rings    
\end{lwcode} 
\caption{Explicit rebindings can be used for solving the constraints arising from the use of \src{sum} and \src{eject multR}. Local resolution grants the expected behaviour.}
\label{fig:ejection-big}
\end{figure}

The inferred types for global functions and relevant sub-expressions are:
$$
\begin{array}{rl}
\src{sum} : & \{~ \src{zero} : \alpha; ~ \src{plus} : \alpha \rightarrow \alpha \rightarrow \alpha ~\} \Rightarrow \alpha ~ \src{list} \rightarrow \alpha
\\
\src{multR} : & \{~ \src{times} : \alpha \rightarrow \beta \rightarrow \beta; ~ \src{one} : \beta ~|~ \kindannot[\rowkind]{\gamma} ~\} \rightarrow \alpha ~ \src{list} \rightarrow \beta
\\
\src{eject multR} : & \{~ \src{?times} : \alpha \rightarrow \beta \rightarrow \beta; ~ \src{?one} : \beta ~\} \Rightarrow \alpha ~ \src{list} \rightarrow \beta
\\
\src{map (eject multR)} : & \alpha ~ \src{list} ~ \src{list} \rightarrow \beta ~ \src{list}
\\
\src{sum (map ...)} : & \{~ \src{zero} : \beta; ~ \src{plus} : \beta \rightarrow \beta \rightarrow \beta ~\} \Rightarrow \beta 
\\
\src{f} : & \{~ \src{times} : \alpha \rightarrow \beta \rightarrow \beta; ~ \src{one} : \beta; ~ \src{zero} : \beta; 
\\
& \src{plus} : \beta \rightarrow \beta \rightarrow \beta ~|~ \kindannot[\rowkind]{\gamma} ~\} \rightarrow \beta
\\
\src{msums} : & \{~ \src{times} : \alpha \rightarrow \beta \rightarrow \beta; ~ \src{one} : \beta; ~ \src{zero} : \beta;  
\\
& \src{plus} : \beta \rightarrow \beta \rightarrow \beta ~|~ \kindannot[\rowkind]{\gamma} ~\} ~ \src{list} \rightarrow \alpha ~ \src{list} ~ \src{list} \rightarrow \beta
\\
\end{array}
$$

Sub-expression $\src{eject multR}$ has constraints in the form of implicit parameters because record fields \src{times} and \src{one} are not overloadable symbols in this scope.
Resolution is greedy in our system and may occur anytime, hence such constraints are immediately solved by local let-bindings \src{times} and \src{one}, \chreplaced{and the type inferred for subexpression \src{map (eject multR)} has the empty constraint set.}
{hence the empty constraint set in the type inferred for subexpression \src{map (eject multR)}.}
Two more constraints (\src{zero} and \src{plus}) arise as soon as the function \src{sum} is called; these are solved immediately by the two homonymous local instances, hence the final type inferred for the whole f function does not contain constraints.

We can get rid of those inner bindings in function \src{msums} in Figure \ref{fig:ejection-big} by exploiting what injection does - binding record fields to variables under the same name and introducing a new lambda over a record parameter. That is exactly what function \src{f} does manually. 
The function \src{msums} could therefore be written as:
\begin{lwcode}
let msums rings list =
    map (inject sum (map (eject multR) list) rings
\end{lwcode}

The types inferred for the most relevant sub-expressions, from the inner-most to the outer-most, are:
$$
\begin{array}{rl}
\src{eject multR} : & \{~ \src{?times} : \alpha \rightarrow \beta \rightarrow \beta; ~ \src{?one} : \beta ~\} \Rightarrow \alpha ~ \src{list} \rightarrow \beta
\\
\src{map (eject multR)} : & \{~ \src{?times} : \alpha \rightarrow \beta \rightarrow \beta; 
\\
& \src{?one} : \beta ~\} \Rightarrow \alpha ~ \src{list} ~ \src{list} \rightarrow \beta ~ \src{list}
\\
\src{sum (map ...)} : & \{~ \src{?times} : \alpha \rightarrow \beta \rightarrow \beta; ~ \src{?one} : \beta; 
\\
& \src{zero} : \beta; ~ \src{plus} : \beta \rightarrow \beta \rightarrow \beta ~\} \Rightarrow \beta
\\
\src{inject sum ...} : & \{~ \src{?times} : \alpha \rightarrow \beta \rightarrow \beta; ~ \src{?one} : \beta; 
\\
& \src{zero} : \beta; ~ \src{plus} : \beta \rightarrow \beta \rightarrow \beta ~|~ \kindannot[\rowkind]{\gamma} ~\} \rightarrow \beta ~ \src{list}
\\
\src{msums} : & \{~ \src{times} : \alpha \rightarrow \beta \rightarrow \beta; ~ \src{one} : \beta; ~ \src{zero} : \beta;  
\\
& \src{plus} : \beta \rightarrow \beta \rightarrow \beta ~|~ \kindannot[\rowkind]{\gamma} ~\} ~ \src{list} \rightarrow \alpha ~ \src{list} ~ \src{list} \rightarrow \beta
\\
\end{array}
$$

Since there are no local instances, constraints arising from sub-expression \src{sum (map (eject multR) list)} include the ejected record fields \src{times} and \src{one} as well as the overloaded symbols \src{zero} and \src{plus}.
Injecting all of them produces the record argument required by the outer-most call to \src{map}.
As it turns out, the combined use of ejection and injection at different levels of nesting allows for advanced forms of constraint manipulation.

\subsection{Restricted injection and ejection}
\label{sec:restricted-ject}

Injection and ejection can be restricted over a certain subset of symbols, allowing the programmer to inject a subset of constraints or to eject part of a record.
Restricted syntax is $\ejectop ~ x_1 ~ .. ~ x_n ~ \src{in} ~ e$ and $\injectop ~ x_1 ~ .. ~ x_n ~ \src{in} ~ e$, where $e$ is an expression.

\begin{lwcode}
overload pretty : 'a -> string
         (+) : 'a -> 'a -> 'a

let rec flatten r l =
    match l with
    | []      -> r.empty
    | [x]     -> r.sep + (pretty x)
    | x :: xs -> (pretty x) + r.sep + (flatten xs)

let over (+) = (^)   // string append instance
let empty = ""
let s = (inject pretty in 
            eject empty in flatten)
                { pretty = sprintf "%d" }
                { sep = ", " }
                [1; 2; 3]    // s : string = "1, 2, 3"
\end{lwcode}


The nested ejection is $\src{eject empty in flatten} : \{~ \src{?empty} : \alpha; ~ \src{pretty} : \beta \rightarrow \alpha; ~ \src{(+)} : \alpha \rightarrow \alpha \rightarrow \alpha ~\} \Rightarrow \{~ \src{sep} : \alpha ~|~ \kindannot[\rowkind]{\gamma} ~\} \rightarrow \beta ~ \src{list} \rightarrow \alpha$.
The enclosing restricted injection has type $\src{inject pretty in ..} : \{~ \src{?empty} : \alpha; ~ \src{(+)} : \alpha \rightarrow \alpha \rightarrow \alpha ~\} \Rightarrow \{~ \src{pretty} : \beta \rightarrow \alpha ~|~ \kindannot[\rowkind]{\delta} ~\} \rightarrow \{~ \src{sep} : \alpha ~|~ \kindannot[\rowkind]{\gamma} ~\} \rightarrow \beta ~ \src{list} \rightarrow \alpha$.
The ejected symbol \src{empty} has become an implicit constraint \src{?empty} because no overload declaration exists for it in the scope.
It is solved by the let-binding of \src{empty}, and \src{(+)} by the overload instance for strings, eventually computing a string ground value \src{s}.

The restricted injection in the example above adds a new arrow whose domain is the record type $\{~ \src{pretty} : \beta \rightarrow \alpha ~|~ \kindannot[\rowkind]{\delta} ~\}$, instead of merging it into the existing record parameter $\{~ \src{sep} : \alpha  ~|~ \kindannot[\rowkind]{\gamma} ~\}$.
In general, each \injectop adds a new record parameter, each with its own fresh row tail, thus multiple restricted injections are not equivalent to a single restricted injection with multiple identifiers.
However, a full injection is equivalent to a restricted injection with the whole identifier set.
$$
\begin{array}{rcl}
e & : & \{~ x : \tau_1; ~ y : \tau_2 ~\} \Rightarrow \tau_3
\\
\src{inject} ~ e & : & \{~ x : \tau_1; ~ y : \tau_2  ~|~ \kindannot[\rowkind]{\alpha} ~\} \rightarrow \tau_3
\\
\src{inject x in} ~ e & : & \{~ y : \tau_2 ~\} \Rightarrow \{~ x : \tau_1  ~|~ \kindannot[\rowkind]{\alpha} ~\} \rightarrow \tau_3
\\
\src{inject x y in} ~ e & : & \{~ x : \tau_1; ~ y : \tau_2  ~|~ \kindannot[\rowkind]{\alpha} ~\} \rightarrow \tau_3
\\
\src{inject x in inject y in} ~ e & : & \{~ x : \tau_1  ~|~ \kindannot[\rowkind]{\alpha} ~\} \rightarrow \{~ y : \tau_2  ~|~ \kindannot[\rowkind]{\beta}  ~\} \rightarrow \tau_3
\end{array}
$$

Note that constraints can be empty after injection, though records cannot be empty after ejection.

As far as ejection is concerned, since type constraints behave like set-like structures, ejecting a whole record or part of it multiple times does not change the result.
$$
\begin{array}{rcl}
e & : & \{~ x : \tau_1; ~ y : \tau_2 ~|~ \kindannot[\rowkind]{\alpha} ~\} \rightarrow \tau_3
\\
\src{eject} ~ e & : & \{~ x : \tau_1; ~ y : \tau_2  ~\} \Rightarrow \tau_3
\\
\src{eject x in} ~ e & : & \{~ x : \tau_1 ~\} \Rightarrow \{~ y : \tau_2 ~|~ \kindannot[\rowkind]{\alpha} ~\} \rightarrow \tau_3
\\
\src{eject x y in} ~ e & : & \{~ x : \tau_1; ~ y : \tau_2  ~\} \Rightarrow \tau_3
\\
\src{eject x in eject y in} ~ e & : & \{~ x : \tau_1; ~ y : \tau_2  ~\} \Rightarrow \tau_3
\end{array}
$$

Restricted ejection does not consume whole record arguments, but subtyping in extensible records obviously does not allow application of a record argument with \emph{fewer} fields than required.
This complicates things.
Consider the following:
\begin{lwcode}
// sample restricted ejection
let g = eject x in fun a -> a.x + a.y
\end{lwcode}
where $\src{g} : \{~ x : \src{int} ~\} \Rightarrow \{~ y : \src{int} ~|~ \kindannot[\rowkind]{\alpha} ~\} \rightarrow \src{int}$ is the inferred type.
Scaling down the translation pattern used in Figure \ref{fig:eject-translation-example} to a single symbol, that would translate into an ill-formed record application, where a supertype of the record type $\{~ x : \src{int}; ~ y : \src{int} ~|~ \kindannot[\rowkind]{\alpha} ~\}$ expected by the lambda is passed:
\begin{lwcode}
// this translation is wrong!
let g = fun x -> (fun a -> a.x + a.y) { x = x }
\end{lwcode}

Restricted ejection requires special treatment: in order to produce well-formed translated code, full records must always be passed, thus non-ejected fields must be abstracted.
This can be achieved through ejecting the whole record and re-injecting the unwanted symbols.
\begin{lwcode}
// restricted eject must eject all and then inject the rest
let g = inject y in eject fun a -> a.x + a.y
\end{lwcode}
This has the same type  $\src{g} : \{~ x : \src{int} ~\} \Rightarrow \{~ y : \src{int} ~|~ \kindannot[\rowkind]{\alpha} ~\} \rightarrow \src{int}$ mentioned above and expected by the restricted ejection; plus it translates into a well-formed expression:
\begin{lwcode}
let g = fun x ->
            fun r ->
                let y = r.y
                in (fun a -> a.x + a.y) { x = x; y = y }
\end{lwcode}

The innermost record application is added by full ejection; the abstraction of \src{r} and binding of \src{y} comes from the restricted injection; finally, the outermost abstraction of symbol \src{x} is added at generalization time when the remaining constraint $\{~ x: \src{int} ~\}$ is translated according to the ordinary dictionary passing.

\section{Related Work}
\label{sec:related}

The design space we are exploring is not novel.
What we call injection is referred to as explicit dictionary application in \cite{winant2018coherent} and primarily serves as a manual resolution system for Haskell type constraints.
A similar system \cite{Devriese:2011:BST:2034773.2034796} has been proposed for the theorem prover Adga.
First-class type classes for Haskell \cite{Sozeau:2008:FTC:1459784.1459810} is an alternative take on explicit dictionary passing and again equivalent to injection.
The same applies to \cite{dijkstra2005making} which adopts implicits rather than type classes.
Named instances for Haskell \cite{kahl2001named} again propose a solution to the manual disambiguation problem.
The goal of all these systems is to provide an explicit dictionary-passing mechanism specifically for Haskell, which is based on type classes and heavyweight record types.
Our proposal is more general as it is based on fine-grained overloading and extensible records.
Moreover, what we call ejection seems uncovered in the literature.
The combined use of injection and ejection, especially in the restricted form discussed in Section \ref{sec:restricted-ject}, allows the programmer to reuse code across two orthogonal dispatching mechanisms in unprecedented ways.

\subsection{Local overloading, local instances and coherence.}

Local instances are crucial for our system and are worth a comparison, despite not being the main focus of the paper.
The literature on this topic is mostly dedicated to type classes à la Haskell, while our fine-grained overloading system features scoped instances that could be shadowed like in \cite{Camarao:1999:TIO:646189.683411}.
Constraint satisfiability is granted by \cite{Camarao:2004:CSO:1013963.1013974} in this case.
The problem of coherence \cite{jones1993coherence} is known with local instances \cite{bottu2019coherence} and implicits \cite{schrijvers2019cochis}.
Our approach can probably be encoded in the more general framework by \cite{stuckey2005theory}.
Particular attention should be given to polymorphic recursion and value-restriction à la ML.
The goal of this paper is not to present a full-fledged overloading system, though, but a simplified ground for exploring injection and ejection.

Our system also features local overloading in order to let ejection work in any scope.
This is not a common feature due to its complications.
Closed-scope type classes \cite{duggan2002open} address this problem for Haskell, but its heavyweight approach makes things complicated.
Our idea is based on a simple trick: converting overloaded symbols into implicit parameters when the symbol escapes the scope of its principal type declaration.
This produces dynamic scoping with strong static types like in \cite{Lewis:2000:IPD:325694.325708}.



\subsection{Scala.}

Scala traits and implicits are powerful enough to encode type classes as a pattern \cite{Oliveira:2010:TCO:1932682.1869489}.
Such a pattern is based on named instance objects acting as dictionaries that can be passed either automatically by the type-driven resolution algorithm or explicitly by the user.
The automatic behavior corresponds to the automatic resolution of overload or implicit constraints in our system, while the manual behavior is basically equivalent to injection.
Ejection is not supported by Scala: there is no way to transform object parameters into implicits.
Also, Scala uses a heavyweight approach to declare implicits, while our question-marked identifiers may occur anywhere in an expression without prior declaration, as in \cite{Lewis:2000:IPD:325694.325708}.
This allows supports the escaping of overloadable identifiers in our system, as shown in Section \ref{sec:interactions-implicits}.

\subsection{Adga and Coq.}

In the world of theorem provers, two major proposals resemble injection: first-class type classes for Coq  \cite{Sozeau:2008:FTC:1459784.1459810} and instance arguments for Agda \cite{Devriese:2011:BST:2034773.2034796}.
With a difference though: in our system records are not heavyweight datatypes and constraints are fine-grained, allowing the programmer to operate on single constraints or record fields.
Furthermore, ejection is novel and the combination of injection and ejection as a reversible programming pattern for general-purpose functional languages has never been explored, to the best of our knowledge.

As far as Agda is concerned, it is interesting to point out that instance arguments are exactly records because dictionary passing is implemented with records.
This is opposed to Haskell or even our own system proposed in this paper, where dictionary passing is implemented through nested lambdas and curried applications.


\subsection{Implicit Calculus.}
Injection and ejection may be formally encoded in implicit calculus \cite{oliveira2012implicit}, though due to space limitations, we provide only a sketch of it.
The bare constructs operating on whole constraint sets or records could be described as follows.

$$
\begin{array}{rclcl}
\src{eject} & : & (\alpha \rightarrow \beta) \rightarrow \alpha \Rightarrow \beta
& = & \lambda x : \alpha \rightarrow \beta.~ \lambda ?\alpha.~ x ~ ?\alpha
\\
\src{inject} & : & (\alpha \Rightarrow \beta) \rightarrow (\alpha \rightarrow \beta)
& = & \lambda c : \alpha \Rightarrow \beta.~ \lambda x : \alpha.~ c ~ \src{with} ~ x
\end{array}
$$

Restricted ejection is harder to encode in implicit calculus as it manipulates parts of records, which are not directly supported.
Adding a row-type system to implicit calculus for expressing fine-grained record field manipulation and encoding restricted injection and ejection may be worth inspecting in a future paper.

\section{Type System}
\label{sec:type-system}

In this section we delve into the formalization of our system, providing syntax-directed type rules, a type inference algorithm and a unification algorithm as well as the relevant proofs of the correctness of the translation performed by type rules.

\begin{bnf}{Syntax of terms. Types $\tau$ are kind-annotated and kinds $\kappa$ are checked syntactically. Judgements include constraints $\pi$ and a translated expression $e^*$}
\label{tab:ast}
e^* & ::= & x ~|~ o ~|~ ?x ~|~ \lambda x. e ~|~ e_1 ~ e_2 ~|~ \src{let} ~ x = e_1 ~ \src{in} ~ e_2 ~|~ e.l ~|~ \{ ~ \many{l} = \many{e} ~\} 
& \termdescr{basic expressions}
\\
e   & ::= & e^* ~|~ \src{overload} ~ o : \tau ~ \src{in} ~ e ~|~ \src{let} ~ \src{over} ~ o = e_1 ~ \src{in} ~ e_2 
& \termdescr{expressions} 
\\
    & |   & \ejectop ~ e ~|~ \injectop ~ e ~|~ \ejectop ~ \overline{x} ~ \src{in} ~ e ~|~ \injectop ~ \overline{x} ~ \src{in} ~ e
& 
\\
\kappa & ::= & \star ~|~ \rowkind ~|~ \kappa_1 \rightarrow \kappa_2
& \termdescr{kinds} 
\\
\tau & ::= &  c^\kappa ~|~ \alpha^\kappa ~|~ \tau_1^{\kappa_1 \rightarrow \kappa_2} \tau_2^{\kappa_1}
& \termdescr{types} 
\\
\pi & ::= & \varnothing ~|~ \pi \cup \{~ x : \tau ~\} & \termdescr{constraints}
\\
\sigma & ::= & \forall \many{\alpha}. \pi \Rightarrow \tau & \termdescr{type schemes}
\\
\Gamma & ::= & \varnothing ~|~ \Gamma,~ x : \sigma & \termdescr{environments}
\\
\theta & ::= & \varnothing ~|~ [ \many{\alpha} \mapsto \many{\tau} ] & \termdescr{substitutions}
\\
\multicolumn{3}{l}{x, o, l} & \termdescr{identifiers}
\end{bnf}

Table \ref{tab:ast} defines the syntax of terms. 
Expressions $e^*$ are a subset of expressions $e$ and represent the target of the translation in type rules. Constraints $\pi$ are possibly empty sets of bindings $\{ x_1 : \tau_1 ~ .. ~ x_n : \tau_n \}$ for $n \geq 0$, where each symbol $x$ can either be a overloaded symbol $o$ or an implicit name $?x$ beginning with a question mark.

Kinds can be star, row and arrows.
Types are kind-annotated and the annotation in the type application $\tau_1^{\kappa_1 \rightarrow \kappa_2} \tau_2^{\kappa_1}$ grants the correctness of the application in a syntactic way. For example, let $\src{list}^{\star \rightarrow \star}$ be the usual type constructor for lists, then the type application $\src{list}^{\star \rightarrow \star} \src{int}^\star$ is kind-correct because the kind of the right-hand matches the domain of the kind-arrow in the left hand. The kind of the whole type application is $\star$, as appears in the codomain of the kind-arrow in the left hand\footnote{Please note that in this section we adopt a right-handed notation for the type application, i.e. the type argument appears on the right. This is on a par with the Haskell syntax and most of the literature, as it allows curried application of type arguments in a natural way. The sample code snippets in the previous sections of this paper, though, adopt an ML style where the type application is left-handed.}.

Types are powerful enough to encode arrows and rows without introducing additional terms in the syntax of $\tau$. We can define built-in types, arrow types and row manipulators just as a plain series of type constructors matching the form $c^\kappa$ in the syntax of types\footnote{To ease readability, we sometimes replace the superscript annotation for kinds as in $\kindannot[\kappa]{\tau}$ with a double-colon notation $\tau :: \kappa$.}.
$$
\begin{array}{lcll}
\src{int} & :: & \star & \termdescr{integers}
\\
(\rightarrow) & :: & \star \rightarrow \star \rightarrow \star & \termdescr{arrow}
\\
\rowempty & :: & \rowkind & \termdescr{empty row}
\\
\rowext & :: & \star \rightarrow \rowkind \rightarrow \rowkind & \termdescr{extend row with label $l$}
\\
\src{record} & :: & \rowkind \rightarrow \star & \termdescr{record}
\\
\end{array}
$$

For writing rows as a sequence of bindings plus a tail, we use an alias for $\rowext$:
$$
\begin{array}{lcl}
\llparenthesis ~ l : \tau ~ | ~ r ~ \rrparenthesis 
& \equiv &
\rowext[l] ~ \tau^\star ~ r^\rowkind
\end{array}
$$

Where $r$ is a syntactic placeholder for any type expression of kind $\rowkind$. This yields to the following shortcuts for open rows and records for $n \geq 1$:
$$
\begin{array}{lcl}
\llparenthesis ~ l_1 : \tau_1 ~ .. ~ l_n : \tau_n ~ | ~ \alpha ~ \rrparenthesis 
& \equiv &
\rowext[l_1] ~ \tau_1 ~ (.. ~ (\rowext[l_n] ~ \tau_n ~ \alpha))
\\
\{ ~ l_1 : \tau_1 ~ .. ~ l_n : \tau_n ~ | ~ \alpha ~ \}
& \equiv &
\src{record} ~ \llparenthesis ~ l_1 : \tau_1 ~ .. ~ l_n : \tau_n ~ | ~ \alpha ~ \rrparenthesis
\end{array}
$$

Closed rows or records are just open rows or records with the empty row $\rowempty$ as tail:
$$
\begin{array}{lcl}
\llparenthesis ~ l_1 : \tau_1 ~ .. ~ l_n : \tau_n ~ \rrparenthesis
& \equiv &
\llparenthesis ~ l_1 : \tau_1 ~ .. ~ l_n : \tau_n ~ | ~ \rowempty ~ \rrparenthesis
\\
\{ ~ l_1 : \tau_1 ~ .. ~ l_n : \tau_n ~ \}
& \equiv &
\{ ~ l_1 : \tau_1 ~ .. ~ l_n : \tau_n ~ | ~ \rowempty ~ \}
\end{array}
$$

With this representation, rows become lists of pairs (label, type) constructed syntactically by nesting type applications, in the same way as a list value in ML is a nesting of data constructors.

Substitutions are straightforward: a substitution $\theta$ can either be empty or a series of mappings $[\many{\alpha} \mapsto \many{\tau}]$ from type variables $\alpha$ to types $\tau$.
The application $\substapp{\tau} = \tau'$ is a kind-preserving substitution.
The composition of two substitutions $\theta_1$ and $\theta_2$ is a third substitution $\theta_3 = \theta_1 \cdot \theta_2$ such that $\substapp[\theta_3]{\tau} = \substapp[\theta_1]{\substapp[\theta_2]{\tau}}$.

We calculate \emph{free type variables} through a family of functions $\fvop$ operating over a variety of terms and producing a set of type variables without kind annotation. Supported terms are: types $\tau$, constraints $\pi$, constrained types $\pi \Rightarrow \tau$, schemes $\sigma$ and environments $\Gamma$. We give as example the calculation of the free type variables of a constrained type: $\fvapp{\{ x_1 : \beta^\star; ~x_2 : \src{int} \} \Rightarrow \alpha^\star \rightarrow \gamma^\star \rightarrow \{~ y_1 : \src{int} ~|~ \delta^\rowkind ~\}} = \{ \alpha, \beta, \gamma, \delta ~\}$.

\subsection{Instances}

Type signatures for overloaded symbols have form $\tau$, whereas instances are bound to type schemes $\sigma$ in the environment, like ordinary let-bindings.

\begin{definition}[Instance Relation]
\label{def:instance-relation}
Given a type $\tau_0$ and a type scheme $\sigma_1 = \forall \many{\alpha}. \pi_1 \Rightarrow \tau_1$, we say that $\sigma_1$ is an instance of $\tau_0$, written $\sigma_1 \isinstanceofop \tau_0$, \emph{iff} a substitution $\theta$ exists such that $\substapp{\tau_0} = \tau_1$, under the assumption that $\{~ \fvapp{\sigma_1} ~\} \cap \{~ \fvapp{\tau_0} ~\} = \varnothing$.
\end{definition}

The latter assumption is satisfiable in practice by simply refreshing the type signature $\tau_0$ of an overloadable symbol.
Moreover, an instance relation between two types, namely $\tau_1 \isinstanceofop \tau_0$, easily derives from the definition above by considering a type $\tau_1$ as a trivial form of type scheme $\forall \varnothing. ~ \varnothing \Rightarrow \tau_1$.

Instances are bound to $\Gamma$ with unique names of form $o^k$, with $k \geq 0$, whereas the original type signature for an overloadable symbol $o$ is bound to the environment with the name $o^0$.
This makes instances always distinct from the original overloadable symbols.
The number $k$ in the superscript of instance names can be calculated through some hashing function over the type scheme inferred for the instance.

\begin{definition}[Hashing of Instance Suffixes]
\label{def:hash}
Let $\hashop : \sigma \rightarrow \mathbb{N}^*$ be a deterministic hash function for type schemes $\sigma \equiv \forall \overline{\alpha}.~ \pi \Rightarrow \tau$.
\end{definition}

We do not give an implementation for $\hashop$, but its result is granted to be greater than 0.
This produces scoped instances, where shadowing may occur only when instances share the same name and the same type scheme.

\begin{lwcode}
overload add : 'a -> 'a -> 'a
let twice x = add x x 

let over add x y = x + y // int -> int -> int
let over add x y = x * y // shadows the instance above
let sixteen = twice 4    // solves with multiplication
\end{lwcode}

This implies there is no global uniqueness of instances in our system, unlike in the presence of type classes à la Haskell.

\begin{definition}[Type Distance]
\label{def:distance}
Given a type signature $\tau$ and an instance whose type scheme is $\sigma$ for which the instance relation $\sigma \isinstanceofop \tau$ holds for some substitution $\theta = [~ \alpha_1 \mapsto \tau_1 ~ .. ~ \alpha_n \mapsto \tau_n ~]$, we define the \emph{type distance} as:
$$
\distapp{\sigma}{\tau} = \sum^n_{i = 1} \rankapp{\tau_i}
$$
where $\rankop : \tau \rightarrow \mathbb{N}$ is a ranking function recursively defined over kind-unannotated types:
$$
\begin{array}{rcl}
\rankapp{\alpha} & = & 0
\\
\rankapp{c} & = & 1
\\
\rankapp{\tau_1 ~ \tau_2} & = & \rankapp{\tau_1} + \rankapp{\tau_2}
\end{array}
$$
\end{definition}

\chdeleted{For example, given $\tau$ and $\sigma$, the relation $\tau \isinstanceofop \sigma$ as well as the distance $\distapp{\tau}{\sigma}$ does make sense - or even the other way round, swapping $\tau$ with $\sigma$.}
The type ranking function $\rankop$ basically considers \emph{variable-to-variable} substitutions as zero-distant, while substitutions from variables to constructors produce a distance equal to 1. This makes for example $\distapp{\src{int}}{\alpha} < \distapp{\src{int} \rightarrow \src{int}}{\alpha}$
or also $\distapp{\src{int} \rightarrow \src{int}}{\alpha \rightarrow \alpha} < \distapp{(\src{int}  \rightarrow \src{int}) \rightarrow \src{int}}{\alpha \rightarrow \alpha}$.

\subsection{Constraints Resolution}

\chadded{Before proceeding with the formalization of the constraint resolution subsystem, we introduce a few basic definitions.
A \emph{best-fitting instance} is the least type-distant among the candidate instances capable of solving a constraint.}

\begin{definition}[Best Fitting Instance]
\label{def:best-fit}
Given an environment $\Gamma$ and a solvable constraint for an overloaded symbol $o : \tau_0$, we define as \emph{best fitting} an instance $\varover{o} : \sigma \in \Gamma$, with $k \geq 1$, whose type distance from $\tau_0$ is minimum compared to other candidates. That is, let $d = \distapp{\sigma}{\tau_0}$, then we say that $\varover{o} : \sigma$ is a \emph{best fitting instance} \emph{iff} $\nexists ~ \varover[k']{o} : \sigma' \in \Gamma ~|~ \distapp{\sigma'}{\tau_0} < d$.
\end{definition}

\chadded{Implicits are treated differently: they can be solved both by overloaded instances \emph{and} by plain let-bindings, depending on whether the symbol, i.e. the implicit identifier, is being overloaded or not in the environment where resolution takes place.}

\begin{definition}[Fitting Binding]
\label{def:fitting-binding}
Given an environment $\Gamma$ and a solvable constraint for an implicit parameter $?x : \tau_0$, we define a \emph{fitting binding} depending on whether the identifier $x$ is overloaded in $\Gamma$ or not. If $\varover[0]{x} \in \Gamma$, a fitting binding is just the best fitting instance introduced by Definition \ref{def:best-fit}.
Otherwise, if $x$ is not overloaded, a fitting binding is any simple binding $x : \sigma \in \Gamma$ such that $\sigma \isinstanceofop \tau_0$, for any type distance $\distapp{\sigma}{\tau_0}$.
\end{definition}

\chadded{Notably, the behaviour above depends on whether the identifier $x$ is overloaded \emph{in the context where resolution takes place}, not where the implicit $?x$ is originally introduced.
This implies that an implicit can be solved lately by a symbol that is not yet overloadable in the original scope.
Consider an example similar to that introduced in Section \ref{sec:interactions-implicits}:}

\begin{lwcode}
// twice : { ?add : 'a -> 'a -> 'b } => 'a -> 'b
let twice x = ?add x x // constraint contains implicit

// symbol add is overloaded afterwards
overload add : 'a -> 'a -> 'a 
let over add = int_plus_int

// four : int
let four = twice 2 // overloaded add solves the implicit
\end{lwcode}

\chadded{This mechanism has another subtle implication: when an implicit is used in a context where that symbol is \emph{already} overloaded, it can virtually be solved immediately by an instance.}

\begin{lwcode}
// symbol add is already overloaded 
overload add : 'a -> 'a -> 'a 
let over add = int_plus_int

// four : int
let four = ?add 2 2 // instance solves implicit immediately
\end{lwcode}

\chadded{If a best-fitting instance is not found, though, the constraint is kept as an implicit, producing no interaction with the type of the overloaded symbol.
The type of \src{four} would therefore be $\{~ \src{?add} : \alpha \rightarrow \alpha \rightarrow \beta ~\} \Rightarrow \beta$.
In other words, implicit constraints are never converted into overloading constraints and preserve their original nature until solved.}

\medskip
Our system performs constraint \emph{simplification} as in \citep{Jones97typeclasses} and similar systems in the literature.
Constraints can be \emph{simplified} in two ways: by \emph{compacting} redundant constraints and by \emph{solving} remaining constraints.

\begin{definition}[Constraints Compaction]
\label{def:compact-constraints}
Let $\compactop : \pi \times e^* \rightarrow \pi \times e^*$ be a function for compacting redundant constraints and performing the proper translation. It is defined recursively by cases:
$$
\begin{array}{rcll}
\compactapp{\{~ o_1 : \tau ~\} \cup \pi}{e^*} & = & \compactapp{\pi}{\src{let} ~ o_1 = o_2 ~ \src{in} ~ e^*}
& \ruleannot{iff $\exists (o_2 : \tau) \in \pi$}
\\
\compactapp{\{~ x_1 : \tau_1 ~\} \cup \pi}{e^*} & = & (\{~ x_1 : \tau_1 ~\} \cup \pi_1; e_1^*)
& \ruleannot{otherwise}
\\
& & \mathtt{where}~(\pi_1, e_1^*) = \compactapp{\pi}{e^*}
\\
\compactapp{\varnothing}{e^*} & = & (\varnothing; e^*) & \ruleannot{fix point}
\end{array}
$$
\end{definition}

Compaction takes place only between constraints coming from overloadable symbols, not from implicits. When the same overloaded symbol $o$ appears as prefix of two constraints $o_1$ and $o_2$ having the same type $\tau$, then $o_1$ is removed from the constraint set $\pi$. The translation introduces a let-binding so that previous occurrences of $o_1$ in $e^*$ are aliased to $o_2$ from now on. The compaction algorithm always terminates, as it consumes constraints within the constraint set $\pi$ until the empty set is left.

\chadded{Resolution, on the other hand, is performed by a function that attempts to solve one \chadded{single} constraint at a time and properly discriminates between overloading constraints and implicit constraints, finding out either a best-fitting instance, a fitting binding or nothing.
We first need to introduce the notion of \emph{solvability} of a constraint, sometimes referred to as \emph{satisfiability} in the relevant literature.
Simply put, a constraint is considered \emph{unsolvable} when a type variable appearing in it does not appear in the type body.}

\begin{definition}[Unsolvable Constraint]
\label{def:unsolvable-constraint}
Given a constrained type $\pi \Rightarrow \tau$, a constraint $x : \tau' \in \pi$ is \emph{unsolvable} \emph{iff} $\fvapp{\tau} \subset \fvapp{\tau'}$.
\end{definition}

\chadded{Unsolvable constraints are left untouched by our solver:}

\begin{definition}[Constraint Solver]
\label{def:solver}
Let $\csolveop : \Gamma \times X \times \tau \rightarrow \{ \varnothing \} \cup (X \times \sigma)$ be a function that, given an environment $\Gamma$ and a constraint $x : \tau$, attempts at finding the best-fitting instance or the fitting binding in $\Gamma$ depending on whether $x$ is an overloaded identifier or an implicit.
In case $x^0 \in \mathtt{dom}(\Gamma)$, then $x$ is an overloaded identifier $o$, therefore $\csolveop(\Gamma; o : \tau) = (\varover[k]{o} : \sigma)$  where $o^k$, with $k \in \mathbb{N}^*$, is the \emph{best-fitting instance} in $\Gamma$ as of Definition \ref{def:best-fit}.
Otherwise $x$ must be an implicit $?y$ prefixed by a question mark, then $\csolveop(\Gamma; ?y : \tau) = (y : \sigma)$ where $y$ is the \emph{fitting binding} in $\Gamma$ as of Definition \ref{def:fitting-binding}.
When no best-fitting instance or fitting binding is found, \chadded{or when the constraint is \emph{unsolvable} as of Definition \ref{def:unsolvable-constraint}}, then $\csolveop(\Gamma; x : \tau) = \varnothing$.
\end{definition}

\chadded{If unsolvable constraints were processed normally by the solver, any available instance would apply, which is undesirable in most cases.
Consider the following snippet, which reproduces the classic Haskell example known as \src{show (read s)}:}

\begin{lwcode}
overload parse : string -> 'a
overload pretty : 'a -> string

let over parse s = float_of_string // some instance for floats

let pp s = pretty (parse s) // parse should remain unsolved
\end{lwcode}

\chadded{The type inferred for function \src{pp} should be $\{~ \src{pretty} : \alpha \rightarrow \src{string}; \src{parse} : \src{string} \rightarrow \alpha ~\} \Rightarrow \src{string} \rightarrow \src{string}$, although the \src{parse} constraint could be immediately solved by the only available instance, as it represents a best-fit according to Definition \ref{def:best-fit}.
In such case, the resulting type would therefore be $\{~ \src{pretty} : \src{float} \rightarrow \src{string} ~\} \Rightarrow \src{string} \rightarrow \src{string}$, locking the \src{float} type in the constraint of \src{pretty} and becoming monomorphic.
This is most likely unwanted, therefore the two constraints of \src{pp} must be ignored by the solver even if some instances of \src{parse} or \src{pretty} exist.}

\chadded{In our system, unsolvable constraints are not treated as ambiguities rejected by the compiler like in \cite{Jones97typeclasses,jones2000type}, but are rather kept and propagated even though they will never be solved automatically.
Programmers may want to disambiguate later by using injection, which is also a form of manual resolution.
In the following snippet, injection is effectively used for passing a record of functions for parsing and pretty-printing integers.}

\begin{lwcode}
let pp_int = (inject pp) { pretty = string_of_int; 
                           parse = int_of_string }
\end{lwcode}

\chadded{The type variable $\alpha$ is unified to \src{int} thanks to the record application, eventually yielding to the unconstrained type $\src{string} \rightarrow \src{string}$.}



\subsection{Type Rules}

\begin{rules}{Type rules I. Basic language terms, record construction, record field selection and constraint resolution.}
\label{tab:type-rules-1}

\myinferrule[width=20em]{Var}
{
\Gamma(x) = \{~ \many{x} : \many{\tau} ~\} \Rightarrow \tau
}
{
\Gamma \vdash x : \ruleout{\{~ \many{\fresh{x}} : \many{\tau} ~\}}{\tau} \transto x ~ \many{\fresh{x}}
}

\myinferrule[width=20em]{Var-O}
{
\Gamma(\varover[0]{o}) = \tau
}
{
\Gamma \vdash o : \ruleout{\{~ \fresh{o} : \tau ~\}}{\tau} \transto \fresh{o}
}

\myinferrule[width=20em]{Var-I}
{
}
{
\Gamma \vdash ?x : \ruleout{\{~ ?x : \tau ~\}}{\tau} \transto  ?x
}

\myinferrule[width=20em]{Abs}
{
\Gamma, (x : \tau_1) \vdash e : \ruleout{\pi}{\tau_2} \transto e^*
}
{
\Gamma \vdash \lambda x. e : \ruleout{\pi}{\tau_1 \rightarrow \tau_2} \transto \lambda x. e^*
}

\myinferrule[width=20em]{App}
{
\Gamma \vdash e_1 : \ruleout{\pi_1}{\tau_2 \rightarrow \tau} \transto e_1^*
\\
\Gamma \vdash e_2 : \ruleout{\pi_2}{\tau_2} \transto e_2^*
}
{
\Gamma \vdash e_1 ~ e_2 : \ruleout{\pi_1 \cup \pi_2}{\tau} \transto e_1^* ~ e_2^*
}

\myinferrule[width=20em]{Sel}
{
\Gamma \vdash e : \ruleout{\pi}{\{~ l : \tau_1 ~|~ \tau_2^\rowkind ~\}} \transto e^*
}
{
\Gamma \vdash e.l : \ruleout{\pi}{\tau_1} \transto e^*.l
}

\myinferrule[width=20em]{Record}
{
\Gamma \vdash e_i : \ruleout{\pi}{\tau_i} \transto e_i^*
\\
(\forall i \in [1, n])
}
{
\Gamma \vdash \{~ l_1 = e_1 ~..~ l_n = e_n  ~\} : \ruleout{\pi}{\{~ l_1 : \tau_1 ~..~ l_n : \tau_n ~\}} \transto \{~ l_1 = e_1^* ~..~ l_n = e_n^*  ~\}
}

\myinferrule{Let}
{
\Gamma \vdash e_1 : \ruleout{\pi_1}{\tau_1} \transto e_1^*
\\
\pi_1 \equiv \{~ \many{x} : \many{\tau} ~\}
\\
\Gamma, (x : \pi_1 \Rightarrow \tau_1) \vdash e_2 : \ruleout{\pi_2}{\tau_2} \transto e_2^*
}
{
\Gamma \vdash \src{let} ~ x = e_1 ~ \src{in} ~ e_2 : \ruleout{\pi_2}{\tau_2} \transto \src{let} ~ x = \lambda \many{x}. ~ e_1^* ~ \src{in} ~ e_2^*
}

\myinferrule{Solve}
{
\Gamma \vdash e : \ruleout{\pi_0 \cup \{~ x_1 : \tau_1 ~\}}{\tau_0} \transto e_0^*
\\
\compactapp{\pi_0}{e_0^*} = (\pi; e^*)
\\
\csolveop(\Gamma; x_1 : \tau_1) = (y : \sigma_1)
\\
\sigma_1 \equiv \forall \many{\alpha}. ~ \pi_1 \Rightarrow \tau_1'
\\
\pi_1 \equiv \{~ \many{x : \tau} ~\}
}
{
\Gamma \vdash e : \ruleout{\pi \cup \{~ \many{\fresh{x}: \tau} ~\}}{\tau_0} \transto \src{let} ~ x_1 = y ~ \many{\fresh{x}} ~ \src{in} ~ e^*
}

\end{rules}

\begin{rules}{Type rules II. Terms dealing with overloading, ejection and injection.}
\label{tab:type-rules-2}

\myinferrule[width=20em]{Overload}
{
o \notin \operatorname{dom}({\pi})
\\
\Gamma, (\varover[0]{o} : \tau_0) \vdash e : \ruleout{\pi}{\tau} \transto e^*
}
{
\Gamma \vdash \src{overload} ~ o : \tau_0 ~ \src{in} ~ e : \ruleout{\pi}{\tau} \transto e^*
}

\myinferrule[width=25em]{Overload-Esc}
{
o \in \operatorname{dom}({\pi})
\\
\Gamma, (\varover[0]{o} : \tau_0) \vdash e : \ruleout{\pi \cup \{~ \fresh{o} : \tau_0 ~\}}{\tau}
\transto e^*
\\
}
{
\Gamma \vdash \src{overload} ~ o : \tau_0 ~ \src{in} ~ e : \ruleout{\pi \cup \{~ ?o : \tau_0 ~\}}{\tau}
\transto \src{let} ~ \fresh{o} = ?o ~ \src{in} ~ e^*
}

\myinferrule{Let-Over}
{
\Gamma \vdash e_1 : \ruleout{\pi_1}{\tau_1} \transto e_1^*
\\
\Gamma(\varover[0]{o}) = \tau_0
\\
\pi_1 \equiv \{~ \many{x} : \many{\tau} ~\}
\\
\tau_1 \isinstanceofop \tau_0
\\
k = \hashapp{\tau_1}
\\
\Gamma, (\varover[k]{o} : \tau_1) \vdash e_2 : \ruleout{\pi}{\tau_2} \transto e_2^*
}
{
\Gamma \vdash \src{let} ~ \src{over} ~ o = e_1 ~ \src{in} ~ e_2 : \ruleout{\pi}{\tau_2}
\transto \src{let} ~ \varover[k]{o} = \lambda \many{x}. ~ e_1^* ~ \src{in} ~ e_2^*
}

\myinferrule[width=20em]{Eject-All}
{
\Gamma \vdash e : \ruleout{\pi}{\{ ~ \many{o} : \many{\tau}; ~ \many{x} : \many{\tau}' ~|~ \alpha^\rowkind ~\} \rightarrow \tau_0} \transto e^*
\\
\forall o \in \many{o}. ~ \varover[0]{o} \in \operatorname{dom}(\Gamma)
}
{
\Gamma \vdash \ejectop ~ e : \ruleout{\pi \cup \{ ~ \many{o} : \many{\tau}; ~ \many{x} : \many{\tau}' ~\}}{\tau_0} \transto e^* ~ \{~ \many{o} = \many{o};~ \many{x} = ?\many{x};  ~\}
}

\myinferrule[width=20em]{Inject-All}
{
\Gamma \vdash e : \ruleout{\{~ \many{x} : \many{\tau} ~\}}{\tau_1} \transto e_1^*
\\
\Gamma \vdash \injectop ~ \many{x} ~ \src{in} ~ e ~ : \ruleout{\pi_2}{\tau_2} \transto e_2^*
}
{
\Gamma \vdash \injectop ~ e : \ruleout{\pi_2}{\tau_2} \transto e_2^*
}

\myinferrule[width=20em]{Eject-R}
{
\Gamma \vdash \ejectop ~ e : \ruleout{\{~ \many{x} : \many{\tau}; ~ \many{y} : \many{\tau}' ~\}}{\tau_1} \transto e_1^*
\\
\Gamma \vdash \injectop ~ \many{y} ~ \src{in} ~ \ejectop ~ e : \ruleout{\pi_2}{\tau_2} \transto e_2^*
}
{
\Gamma \vdash \ejectop ~ \many{x} ~ \src{in} ~ e : \ruleout{\pi_2}{\tau_2} \transto e_2^*
}

\myinferrule{Inject-R}
{
\Gamma \vdash e : \ruleout{\pi \cup \{~ \many{x} : \many{\tau} ~\}}{\tau} \transto e^*
\\
\ruleannot{$y, \alpha^\rowkind$ fresh}
\\
\{~ \many{x} : \many{\tau} ~\} \equiv \{~ x_1 : \tau_1 ~ .. ~ x_n : \tau_n ~\}
\\
n \geq 1
}
{
\Gamma \vdash \injectop ~ \many{x} ~ \src{in} ~ e
: \ruleout{\pi}{\{ ~ \many{x} : \many{\tau} ~ | ~ \alpha ~ \} \rightarrow \tau}
\\
\transto \lambda y. ~ \src{let} ~ x_1 = y.x_1 ~ \src{in let} ~ .. ~ x_n = y.x_n ~ \src{in} ~ e^*
}
\end{rules}

In Table \ref{tab:type-rules-1} and Table \ref{tab:type-rules-2} syntax-directed type rules are presented.
Judgments are formulas $\Gamma \vdash e : \pi. \tau \transto e^*$ assigning a type $\tau$ and a translation $e^*$ to each input term $e$ in the typing context $\Gamma$.
Constraints $\pi$ represent the current constraint set being collected during the type derivation.
Type rules deal with constrained type $\pi \Rightarrow \tau$: when an unconstrained type $\tau$ appears, it is a short form for $\varnothing \Rightarrow \tau$.
Polymorphism, generalization and instantiation are not mentioned here and will be supported by type inference rules presented in Section \ref{sec:type-inference}.
Environment $\Gamma$, therefore, includes constrained types $\pi \Rightarrow \tau$ as dummy type schemes $\forall \varnothing. \pi \Rightarrow \tau$.

\medskip
\chdeleted{In type rules, identifier names can appear in many forms.}
\chadded{Before proceeding with the description of type rules, an explanation of all identifier formats used is necessary for understanding how implicits and overloaded symbols are transformed when bound to the environment.}

\begin{itemize}
    \item Identifiers of form $o$ stand for overloadable symbols.
    When bound in the environment $\Gamma$ a superscript $k \geq 0$ is added to discriminate instances.
    An implementation can append such superscript as a suffix separated from the base identifier by a special character that never appears at the lexical level within identifier names.
    For example, $o^7$ could be encoded as \texttt{o\$7}, assuming the dollar symbol \texttt{\$} does not belong to variable names.
    Rule \rulename{Overload} binds the principal type annotated by the programmer to the environment and adds the $0$ superscript: $o^0$ is, therefore, a special name reserved for the original principal type and no instance ever happens to have such name.
    Instances bound by rule \rulename{Let-Over} add a unique superscript $k \geq 1$ calculated by a hashing function.
    More on this in Definition \ref{def:hash}.
    Specific instances are denoted with $o^k$, e.g. those found by the constraint solver in rule \rulename{Solve}.

    \item Identifiers of form $?x$ are implicits.
    When converted from an overloadable identifier $o$, like in rule \rulename{Overload-Esc}, they appear as $?o$.
    An implementation can just prefix the question mark character for discriminating implicits from ordinary identifiers, assuming the \texttt{?} is not a legal character for identifiers at the lexical level.

    \item Identifiers of form $x$ stand for any identifier, including overloadable identifiers $o$, instance identifiers $o^k$, implicits $?x$ and normal variable names.
    This is a match-all notation for manipulating identifiers of any form in the type rules.
    
    \item When an identifier $o$, $x$ or an implicit $?x$ appears in a constraint, they are always implicitly suffixed with a number.
    Such suffix is omitted in our formalization for the sake of simplicity.
    This does not have to be confused with the $k$ superscript added for discriminating instances.
    An implementation may treat this as an integer suffix that is separated from the base identifier $o$ by a different separator, e.g. the \texttt{\%} rather than the \texttt{\$} used for instances.
    This is a well-known practice in open-world overloading systems \cite{Wadler:1989:MAP:75277.75283,jones1992theory,Jones97typeclasses,Odersky:1995:SLO:224164.224195} that is necessary for allowing multiple occurrences of the same overloaded identifier within the same expression.
    Each occurrence must produce a distinct constraint that could be solved by a different instance.    

    \item Bolded identifiers of form $\fresh{o}$ represent overloaded identifiers whose name is being refreshed by replacing the numerical suffix with a new number.
    Dictionary passing is responsible for this, and rule \rulename{Var} shows the mechanism in action.
    
    \item Bolded identifiers $\fresh{x}$ stand for identifiers of any form being refreshed due to dictionary passing.

    \item Identifiers of form $l$ are record labels appearing in rules \rulename{Sel} and \rulename{Record}.
    Injection and ejection treat record labels as identifiers either of form $x$ or $o$, depending on what is stored in the environment, as rule \rulename{Eject-All} shows.

    \item Over-lined identifiers, such as $\many{x}$, $\many{o}$ or $?\many{x}$, represent sequences of identifiers. 

\end{itemize}

We now proceed with the description of the most relevant rules in Table \ref{tab:type-rules-1}.
Rule \rulename{Var} is responsible for dictionary passing.
Constraints retrieved from $\Gamma$ when looking up variable $x$ are inherited, and all constraint names are refreshed with new suffixes and added to the current constraint set.
Refreshed names $\many{\fresh{x}}$ are applied in currying to $x$ to pass the dictionary.
Rule \rulename{Var-O} deals with overloadable variable identifiers of form $o$.
These can be discriminated from ordinary variables $x$ by looking up a principal type bound to $o^0$ in $\Gamma$.
If the lookup succeeds, identifier $o$ gets a unique numeric suffix and is inserted into the current constraint set.
This suffix generation is treated as a refreshment of the name and is represented by the bolded $\fresh{o}$.
If the lookup fails, rule \rulename{Var} holds.
Rule \rulename{Var-I} holds when an identifier prefixed with a question mark is encountered.
Implicits do not get a suffix: this implies that multiple occurrences of the same implicit within an expression do not produce distinct constraints, as opposed to what happens with overloaded identifiers.
The two behaviours are explained in Section \ref{sec:interactions-implicits}.
Rules \rulename{Abs}, \rulename{App} and \rulename{Let} are normal Hindley-Milner rules for lambda abstraction, application and let-binding, respectively, extended with type constraints.
Additionally, rule \rulename{Let} lambda abstracts constraints to allow dictionary passing: $\lambda \many{\fresh{x}}$ stands for many nested lambdas.
Rules \rulename{Sel} and \rulename{Record} employ row types and deal with extensible records.
These are taken from the extensible record system with scoped labels described in \cite{Leijen:2009:FTR:1480881.1480891} and adapted to work with type constraints.

Rule \rulename{Solve} deals with constraint resolution and may be used at any time during a type derivation.
It performs constraint compaction before solving, simplifying the constraint set $\pi_0$ and producing a $\pi$ that appears in the rule output.
The rule holds whenever some constraint $x_1 : \tau_1$ is solvable, i.e. when the solver $\csolveop$ finds either a best-fitting instance or a fitting binding $y : \sigma_1$ in the current context $\Gamma$.
The identifier $x_1$ can either be an overloaded identifier or an implicit, thus the resolved $y$ can either be an instance of form $o^k$ or a plain binding.
Since the instance found by the solver may be a constrained function itself, its constraint set $\pi_1$ is inherited and refreshed, hence the bold notation $\fresh{\many{x}}$ standing for all identifiers $\many{x}$ being inherited from $\pi_1$ with a fresh suffix.
The translation produces a \src{let} binding the constraint name $x_1$ to the name $y$. Dictionary passing produces a curried application of refreshed identifiers $\fresh{\many{x}}$ due to constraint inheritance, which reproduces the same behaviour as in the \rulename{Var} rule.

Table \ref{tab:type-rules-2} contains type rules for terms dealing with overloading, injection and ejection.
Rule \rulename{Overload} binds the principal type $\tau_0$ to the reserved name $o^0$.
No occurrence of $o$ must remain in the constraint set $\pi$ after expression $e$ has been derived.
This means that the overloadable symbol $o$ must not escape its scope.
Rule \rulename{Overload-Esc} holds otherwise: when the overloadable symbol $o$ escapes its scope, it gets converted into an implicit $?o$ on the fly.
This happens when a constraint for $o$ remains unsolved, i.e. when the suffixed overloadable identifier represented by the bold $\fresh{o}$ is left in the constraint set.
The translation ensures that such suffixed identifier $\fresh{o}$ is aliased with the implicit $?o$.
Note that the implicit is not suffixed: this implies that multiple occurrences of $o$ having distinct suffixes would not become multiple implicits, but would rather be aliased to the same $?o$, producing a single constraint.
This comes naturally from constraints being a set: multiple occurrences of the same constraint identifier would require to have the same types, otherwise an error must occur.

An implementation featuring type inference would unify the type parts of duplicated constraints.
Take the following example:
\begin{lwcode}
let f x =           // { ?id : 'a -> 'a } => 'a -> 'a
    overload id : 'a -> 'b
    in id (id x)    // { id%1 : 'a -> 'b; 
                    //   id%2 : 'b -> 'c } => 'a -> 'c
\end{lwcode}

The constraints collected for the expression \lstinline{id (id x)} include two occurrences of the locally overloadable function \lstinline{id (id x)}, namely the suffixed names \lstinline{id
They will both escape their scope when typing reaches the \lstinline{overload} construct, as no instances are found.
They would both be converted into one single implicit \lstinline{?id} and their types unified, yielding to the final type of \src{f} shown in the comment.

Rule \rulename{Let-Over} is very similar to rule \rulename{Let}.
Additionally, the type of the instance is checked to be in relation to the original principal type annotated by the programmer for the overloadable symbol.

Rule \rulename{Eject-All} expects expression $e$ to be an arrow type with a record domain.
It basically splits the record fields into two groups: identifiers $o$ that are overloaded in $\Gamma$ and other labels of form $x$ that are not overloaded.
Whether a record label $x$ has a corresponding binding with the same name in $\Gamma$ or not, it is stubbed to an implicit.
The solver can solve implicits with plain let-bindings anyway.
Translation erases the arrow type by introducing the application of a synthetic record that stubs all fields into either overloaded symbols or implicit parameters.

Rule \rulename{Eject-R} actually implements a syntactic sugar of \rulename{Eject-All} and \rulename{Inject-R}, as discussed in Section \ref{sec:restricted-ject}.
The same applies to rule \rulename{Inject-All}.

Rule \rulename{Inject-R} is native instead.
It moves the injected identifiers from the constraints to a new record that becomes the domain of the resulting arrow type.
Translation adds a synthetic lambda abstraction: the lambda parameter $y$ is a record, and inside the lambda body all constraint identifiers $x_1 .. x_n$ are bound to the fields of the record with the same names.

\section{Type Inference}
\label{sec:type-inference}

\begin{rules}{Type inference algorithm $w$ in form of syntax-directed rules I.}
\label{tab:infer-rules-1}

\myinferrule{$w$-Var}
{
\Gamma(x) = \sigma 
\\
\instantiate{\sigma} = \{~ \many{x} : \many{\tau} ~\} \Rightarrow \tau_0
}
{
\Gamma \wdash x : \ruleout{\{~ \many{\fresh{x}} : \many{\tau} ~\}}{\tau_0} \transto x ~ \many{\fresh{x}} \rhd \varnothing
}

\myinferrule{$w$-Var-O}
{
\Gamma(\varover[0]{o}) = \sigma
\\
\sigma \equiv \forall \many{\alpha}. \varnothing \Rightarrow \tau
\\
\instantiate{\sigma} = \varnothing \Rightarrow \tau_0
}
{
\Gamma \wdash o : \ruleout{\{~ \fresh{o} : \tau_0 ~\}}{\tau_0} \transto \fresh{o} \rhd \varnothing
}

\myinferrule[width=20em]{$w$-Var-I}
{
\ruleannot{$\alpha$ fresh}
}
{
\Gamma \wdash ?o : \ruleout{\{~ ?o : \alpha ~\}}{\alpha} \transto ?o \rhd \varnothing
}

\myinferrule{$w$-Solve}
{
\Gamma \wdash e : \infout{\pi_0 \cup \{~ x_1 : \tau_1 ~\}}{\tau_0}{e_0^*}{\theta_1}
\\
\compactapp{\pi_0}{e_0^*} = (\pi; e^*)
\\
\csolveop(\Gamma; x_1 : \tau_1) = (y : \sigma_1)
\\
\instantiate{\sigma_1} = \{~ \many{x} : \many{\tau} ~\} \Rightarrow \tau'_1
\\
\tau_1 \sim \tau'_1 \rhd \theta_2
\\
\theta_3 = \theta_2 \cdot \theta_1
}
{
\Gamma \wdash : e : \infout{\pi \cup \{~ \many{\fresh{x} : \tau} ~\}}{\tau_0}{\src{let} ~ x_1 = y ~ \many{\fresh{x}} ~ \src{in} ~ e^*}{\theta_3}
}

\myinferrule{$w$-App}
{
\Gamma \wdash e_1 : \infout{\pi_1}{\tau_1}{e_1^*}{\theta_1}
\\
\substapp[\theta_1]{\Gamma} \wdash e_2 : \infout{\substapp[\theta_1]{\pi_2}}{\substapp[\theta_1]{\tau_2}}{e_2^*}{\theta_2}
\\
\substapp[\theta_2]{\tau_1} \sim \tau_2 \rightarrow \alpha \rhd \theta_3
\\
\theta_4 = \theta_3 \cdot \theta_2 \cdot \theta_1
}
{
\Gamma \wdash e_1 ~ e_2 :
    \infout{\substapp[\theta_4]{\pi_1 \cup \pi_2}}
    {\substapp[\theta_4]{\alpha}}
    {e_1^* ~ e_2^*}
    {\theta_4}
}

\myinferrule{$w$-Sel}
{
\Gamma \wdash e : \infout{\pi}{\tau}{e^*}{\theta_1}
\\
\tau \sim \{~ l : \alpha ~|~ \beta ~\} \rhd \theta_2
\\
\theta_3 = \theta_2 \cdot \theta_1
\\
\ruleannot{$\alpha^\star, \beta^\rowkind$ fresh}
}
{
\Gamma \wdash e.l : \infout{\substapp[\theta_3]{\pi}}{\substapp[\theta_3]{\alpha}}{e^*.l}{\theta_3}
}

\end{rules}

\begin{rules}{Type inference algorithm $w$ in form of syntax-directed rules II.}
\label{tab:infer-rules-2}

\myinferrule{$w$-Let}
{
\Gamma \wdash e_1 : \infout{\pi_1}{\tau_1}{e^*_1}{\theta_1}
\\
\pi_1 = \{~ \many{x} : \many{\tau} \}
\\
\many{\alpha} = (\fvapp{\pi_1} \cup \fvapp{\tau_1}) \setminus \fvapp{\Gamma}
\\
\sigma_1 = \forall \many{\alpha}. ~ \pi_1 \Rightarrow \tau_1
\\
\substapp[\theta_1]{\Gamma}, (x : \sigma_1) \wdash e_2 : \infout{\pi_2}{\tau_2}{e_2^*}{\theta_2}
}
{
\Gamma \wdash \src{let} ~ x = e_1 ~ \src{in} ~ e_2 : \infout{\pi_2}{\tau_2}
    {\src{let} ~ x = \lambda \many{x}.~ e^*_1 ~ \src{in} ~ e^*_2}
    {\theta_2}
}

\myinferrule{$w$-Let-Over}
{
\Gamma \wdash e_1 : \infout{\pi_1}{\tau_1}{e^*_1}{\theta_1}
\\
\pi_1 = \{~ \many{x} : \many{\tau} \}
\\
\many{\alpha} = (\fvapp{\pi_1} \cup \fvapp{\tau_1}) \setminus \fvapp{\Gamma}
\\
\sigma_1 = \forall \many{\alpha}. ~ \pi_1 \Rightarrow \tau_1
\\
\sigma_1 \isinstanceofop \Gamma(\varover[0]{o})
\\
\substapp[\theta_1]{\Gamma}, (x : \sigma_1) \wdash e_2 : \infout{\pi_2}{\tau_2}{e_2^*}{\theta_2}
\\
\ruleannot{$k \geq 1$ fresh}
}
{
\Gamma \wdash \src{let over} ~ o = e_1 ~ \src{in} ~ e_2 : \infout{\pi_2}{\tau_2}
    {\src{let} ~ x = \lambda \many{x}.~ e^*_1 ~ \src{in} ~ e^*_2}
    {\theta_2}
}

\myinferrule{$w$-Eject-All}
{
\Gamma \wdash e : \infout{\pi_1}{\tau_1}{e_1^*}{\theta_1}
\\
\tau_1 \sim \{~ \beta ~\} \rightarrow \alpha \rhd \theta_2
\\
\theta_2(\{~ \beta ~\}) = \{~ \many{o} : \many{\tau}; ~ \many{x} : \many{\tau'} ~|~ \gamma^\rowkind ~\}
\\
\forall o \in \many{o}. ~ \varover[0]{o} \in \operatorname{dom}(\Gamma)
\\
\theta_3 = \theta_2 \cdot \theta_1
\\
\ruleannot{$\alpha^\star, \beta^\rowkind$ fresh}
}
{
\Gamma \wdash \ejectop ~ e : \infout{\{~ \many{o} : \many{\tau}; ~ \many{x} : \many{\tau'} ~\}}{\theta_3(\alpha)}{e_1^* ~ \{~ \many{o} = \many{o};~ \many{x} = ?\many{x};  ~\}}{\theta_3}
}

\myinferrule{$w$-Eject-R}
{
\Gamma \wdash \src{inject} ~ \many{x} ~ \src{in eject} ~ e : \infout{\pi}{\tau}{e^*}{\theta}
}
{
\Gamma \wdash \ejectop ~ \many{x} ~ \src{in} ~ e : \infout{\pi}{\tau}{e^*}{\theta}
}

\myinferrule{$w$-Inject-All}
{
\Gamma \wdash e : \infout{\pi_1}{\tau_1}{e_1^*}{\theta_1}
\\
\pi_1 = \{~ \many{x} : \many{\tau} ~\}
\\
\Gamma \wdash \src{inject} ~ \many{x} ~ \src{in} ~ e : \infout{\pi_2}{\tau_2}{e_2^*}{\theta_2}
\\
}
{
\Gamma \wdash \injectop ~ e :\infout{\pi_2}{\tau_2}{e_2^*}{\theta_2}
}

\myinferrule{$w$-Inject-R}
{
\Gamma \wdash e : \infout{\pi}{\tau}{e^*}{\theta_1}
\\
\ruleannot{$y, \kindannot[\rowkind]{\alpha}$ fresh}
}
{
\Gamma \wdash \injectop ~ \many{x} ~ \src{in} ~ e :
    \pi \backslash \{ ~ \many{x} : \many{\tau} ~\}. \{ ~ \many{x} : \many{\tau} ~ | ~ \alpha ~ \} \rightarrow \tau
    \\
    \transto
    \lambda y. ~ \src{let} ~ x_1 = y.x_1 ~ \src{in let} ~ .. ~ x_n = y.x_n ~ \src{in} ~ e^*
    \rhd
    \theta_1
}

\end{rules}

Our type inference algorithm is a function $w : \Gamma \times e \rightarrow \pi \times \tau \times e^* \times \theta$ formulated as a deduction system with syntax-directed inference clauses of form $\Gamma \wdash e : \infout{\pi}{\tau}{e^*}{\theta}$.
\chadded{The environment $\Gamma$ and the expression $e$ can be considered the inputs of the rule, whereas the outputs consist in the constraint set $\pi$, the type $\tau$, the translated expression $e^*$ and the substitution $\theta$.
The formula can be read as a standard Hindley-Milner type inference clause: in a context $\Gamma$, an expression $e$ is inferred to have type $\tau$ under the substitution $\theta$. Additionally, our rules produce the constraint set $\pi$, where overloadable symbols and implicits encountered during typing are collected, and perform translation of the input expression $e$ into a simpler form $e^*$, revealing the dictionary passing mechanism.}

We present a description of all type inference rules in Table \ref{tab:infer-rules-1} and Table \ref{tab:infer-rules-2}.
In Table \ref{tab:infer-rules-1} rules \rulename{$w$-Var} and \rulename{$w$-Var-O} are equivalent to the rules of the respective type in Table \ref{tab:type-rules-1}. 
The instantiation function $\instantiate{\sigma}$ for a type scheme $\sigma \equiv \forall \many{\alpha}. \pi \Rightarrow \tau$ is the typical Hindley-Milner instantiation extended for constrained types, where each $\alpha \in \many{\alpha}$ is substituted with a new fresh type variable $\beta$ both in the constraint set $\pi$ and in the type body $\tau$.

Rule \rulename{$w$-Var-I} introduces a fresh type variable when an implicit is encountered.

Rule \rulename{$w$-Solve} behaves like type rule \rulename{Solve} in Table \ref{tab:type-rules-1}, performing constraints compaction and resolution at any time, as long as the solver finds an instance for a given constraint $x_1$.
Additionally, it performs instantiation of the type scheme $\sigma_1$ of the instance $o^k$ found by the solver, eventually unifying its refreshed type $\tau_1'$ with the constraint original type $\tau_1$.
Since the instance may be a constrained function itself, its constraint set is inherited and dictionary passing is performed in the translation.

Rules \rulename{$w$-App} and \rulename{$w$-Sel} are straightforward and perform unification, the latter being taken from \cite{extensible-records-with-scoped-labels}.

In Table \ref{tab:infer-rules-2} rules \rulename{$w$-Let} and \rulename{$w$-Let-Over} perform generalization.
Function $\fvop$ calculates free type variables of constraints, types or environments according to the usual definition for Hindley-Milner type systems.
Also, abstraction is performed in translation, in the same way as the respective type rules do.

Rule \rulename{$w$-Eject} forces unification of the type of expression $e$ with an arrow type having a record type in the domain.
Such record type is represented by the row type variable $\beta$.
Similarly to the respective type rule in Table \ref{tab:type-rules-2}, we split the record labels to be ejected into two groups: those which are overloaded and those which are not.
Refer to Section \ref{sec:type-system} for more details on this behaviour.

Rule \rulename{$w$-Eject-R}, \rulename{$w$-Inject} and \rulename{$w$-Inject-R} reproduce the same behaviour of the respective type rules.

\subsection{Unification}

\chdeleted{Table \ref{tab:uni1} shows the unification algorithm in the form of syntax-directed rules.
Rules \rulename{U-Const}, \rulename{U-Var} plus both \rulename{U-Var-L} and \rulename{U-Var-R} do not need further explanations.
Rule \rulename{U-App} shows how kind annotations enable a syntactic form of kind checking: since left and right-hand types are defined as $\tau_1^{\kappa_2 \rightarrow \kappa}$ and $\tau_2^{\kappa_2}$, kind unification takes place at a syntactic level, constraining $\kappa_2$ to be both the kind of the domain of $\tau_1$ and the kind of $\tau_2$. In other words, our kind system does not need special rules for unification at the kind level, since unification rules \rulename{U-Var-L} and \rulename{U-Var-R} grant preservation of kinds by allowing unification from and to type variables only when kind-annotations match at the syntactic level.
For dealing with rows we present the same unification rules introduced by \cite{extensible-records-with-scoped-labels}, performing row insertion and swapping in a kind-preserving way.
The main row-rewriting rule is \rulename{U-Row}.
It is algorithmic in nature, as it triggers other row-rewriting rules in its hypothesis to perform unification between row tails.
\rulename{Row-Head} unifies heading elements of rows when labels and types are the same. \rulename{Row-Swap} searches for a proper label $l$ by pushing unwanted ones down the row, hence lifting up $l$ to the head. 
\rulename{Rule-Var} unifies type variables of kind \src{row} with an actual row type term.}









\begin{rules}{Kind-preserving unification algorithm for simple types and row types.}
\label{tab:uni1}

\myinferrule{U-Const}
{
}
{
c^\kappa \sim c^\kappa \rhd \varnothing
}

\myinferrule{U-Var}
{
}
{
\alpha^\kappa \sim \alpha^\kappa \rhd \varnothing
}

\myinferrule{U-Var-L}
{
\alpha \not\in \fvapp{\tau}
}
{
\alpha^\kappa \sim \tau^\kappa \rhd [~ \alpha \mapsto \tau ~]
}

\myinferrule{U-Var-R}
{
\alpha \not\in \fvapp{\tau}
}
{
\tau^\kappa \sim \alpha^\kappa \rhd [~ \alpha \mapsto \tau ~]
}

\myinferrule{U-App}
{
\tau_1 \sim \tau_3 \rhd \theta_1
\\
\substapp[\theta_1]{\tau_2} \sim \substapp[\theta_1]{\tau_4} \rhd \theta_2
}
{
\tau_1^{\kappa_2 \rightarrow \kappa} ~ \tau_2^{\kappa_2}
\sim \tau_3^{\kappa_2 \rightarrow \kappa} ~ \tau_4^{\kappa_2} \rhd \theta_2 \cdot \theta_1
}

\myinferrule[width=25em]{U-Row}
{
s_1 \simeq \llparenthesis ~ l : \tau_2 ~ | ~ s_2 ~ \rrparenthesis \rhd \theta_1
\\
\operatorname{tail}(r_1) \not\in \operatorname{dom}(\theta_1)
\\
\substapp[\theta_1]{\tau_1} \sim \substapp[\theta_1]{\tau_2} \rhd \theta_2
\\
\substapp[\theta_2]{\substapp[\theta_1]{r_1}} \sim \substapp[\theta_2]{\substapp[\theta_1]{s_2}} \rhd \theta_3
}
{
\llparenthesis ~ l : \tau_1 ~ | ~ r_1 ~ \rrparenthesis \sim s_1 \rhd \theta_3 \cdot \theta_2 \cdot \theta_1
}
\end{rules}

\begin{rules}{\chadded{Row-rewriting rules preserving type equality.}}
\label{tab:uni2}

\myinferrule{Row-Head}
{
}
{
\llparenthesis ~ l : \tau ~ | ~ r ~ \rrparenthesis \simeq \llparenthesis ~ l : \tau ~ | ~ r ~ \rrparenthesis \rhd \varnothing
}

\myinferrule{Row-Swap}
{
l_1 \neq l_2
\\
r_1 \simeq \llparenthesis ~ l_1 : \tau_1 ~ | ~ r_2 ~ \rrparenthesis \rhd \theta
}
{
\llparenthesis ~ l_2 : \tau_2 ~ | ~ r_1 ~ \rrparenthesis \simeq \llparenthesis ~ l_1 : \tau_1;~ l_2 : \tau_2 ~ | ~ r_2 ~ \rrparenthesis \rhd \theta
}

\myinferrule{Row-Var}
{
\ruleannot{$\kindannot[\rowkind]{\beta}, \kindannot{\gamma}$ fresh}
}
{
\alpha^\rowkind \simeq \llparenthesis ~ l : \gamma ~ | ~ \beta ~ \rrparenthesis
\rhd [~ \alpha \mapsto \llparenthesis ~ l : \gamma ~ | ~ \beta ~ \rrparenthesis ~]
}
\end{rules}

\chadded{
In our system, substitutions are kind-preserving, meaning that a substitution always maps type variables of a particular kind to types of the same kind. 
A substitution $\theta$ is called a unifier of two types $\tau_1$ and $\tau_2$ only if $\theta(\tau_1) = \theta(\tau_2)$. 
If every other unifier can be written as the composition $\theta' \cdot \theta$, for some substitution $\theta'$, then we call $\theta$ a most general unifier.
To compute the most general unifier $\theta$ for two types $\tau_1$ and $\tau_2$ in the presence of rows, we use the rules provided in Table \ref{tab:uni1}.
The notation $\tau_1 \sim \tau_2 \rhd \theta$ indicates the calculation of the most general unifier $\theta$ for two types $\tau_1$ and $\tau_2$.}

\chadded{The first five rules in Table \ref{tab:uni1} deal with standard kind-preserving unification for simple types, while rule \rulename{U-Row} deals with rows. 
When attempting to unify a row $\llparenthesis~ l : \tau_1 ~|~ r_1 ~\rrparenthesis$ with some other row $s_1$, the latter is rewritten to match the form $\llparenthesis~ l : \tau_2 ~|~ s_2 ~\rrparenthesis$, where $\tau_2$ and $s_2$ are respectively a new type and a new row synthesized by the row-rewriting rules in Table \ref{tab:uni2}.
If the rewriting is successful, the field types $\tau_1$ and $\tau_2$ as well as the tails of the rows $r_1$ and $s_2$ are unified.}

\chadded{Rules for rewriting rows in Table \ref{tab:uni2} present clauses of form $r \simeq \llparenthesis~ l : \tau ~|~ s ~\rrparenthesis \rhd \theta$, asserting that a row $r$ can be rewritten to the form $\llparenthesis~ l : \tau ~|~ s ~\rrparenthesis$ under the substitution $\theta$, with $r$ and $l$ being input parameters while $\tau$ , $s$, and $\theta$ are synthesized. 
Rule \rulename{Row-Head} simply states that two row types having the same head and the same tail are equal. 
Rule \rulename{Row-Swap} states that it is possible to swap the first two fields of a row and still consider it as the same row, as long as the labels of those fields are distinct.
Finally, rule \rulename{Row-Var} unifies a row tail that consists of a type variable $\alpha$ of kind $\rowkind$.
Please note that this rule introduces fresh type variables which might affect the termination of the unification algorithm. 
This is the reason for the side condition in rule \rulename{U-Row}: $\operatorname{tail}(r_1) \not\in \operatorname{dom}(\theta_1)$, where the tail of a row type is intended as the row without its head. 
For a detailed explanation of why this side condition is necessary and for further details, please refer to the paper that first introduced this unification system for extensible rows \cite{extensible-records-with-scoped-labels}.}

\subsection{Correctness of the Translation of Eject and Inject}
\label{sec:soundness}

In this section, we sketch the proofs of correctness of the translation performed by type rules involving \injectop and \ejectop constructs.
Due to the introductory nature of the paper and the preliminary state of the study of the proposed system, we leave a full formalization of the type system to a future work. This would include a proof of the preservation of the semantics, a proof of coherence for our overloading subsystem \cite{jones1993coherence,bottu2019coherence} also in the presence of explicit dictionary passing \cite{winant2018coherent}, declined for our first-class operator \injectop. As well as a proof of coherence for ejection, which is something novel.

For the proofs of soundness of the extensible record subsystem and unification of rows, please refer to the original paper \cite{extensible-records-with-scoped-labels}.

\newcommand{\rowtail}[1][\alpha]{\kindannot[\rowkind]{#1}}

\begin{lemma}[Correctness of Inject-R Translation]
\label{lem:restricted-inject-translation}
Given a type derivation $\Gamma \vdash e_0 : \ruleout{\{ \many{x} : \many{\tau} \} \cup \pi}{\tau_0} \transto e_0^*$
and the derivation of the restricted injection
$\Gamma \vdash \injectop ~ \many{x} ~ \src{in} ~ e_0 : \ruleout{\pi}{\{ \many{x} : \many{\tau} ~|~ \rowtail \} \rightarrow \tau_0} \transto e^*$, 
then a derivation $\Gamma \vdash e^* : \ruleout{\varnothing}{\{ \many{x} : \many{\tau} ~|~ \rowtail \} \rightarrow \tau_0} \transto e^*$ exists.

\proof (Sketched)
Translation $e^* = \lambda y. ~\src{let} ~ x_1 = r.x_1 ~ .. ~ \src{let} ~ x_n = r.x_n ~ \src{in} ~ e^*_0$ comes from rule \rulename{Inject-R}.
The type of the new lambda argument $y$ is unified to the record type $\{ \many{x} : \many{\tau} ~|~ \rowtail \}$ because of the internal let-bindings and rule \rulename{Sel}.
Identifiers $x_1 ~ .. ~ x_n$ must appear in $e^*_0$, introduced by dictionary passing as of rule \rulename{Var}, because they belong to the original constraint set $\{ \many{x} : \many{\tau} \} \cup \pi$ of $e_0$ by hypothesis.
Therefore, rule \rulename{Let} makes the derivation $\Gamma, \many{x} : \many{\tau} \vdash e^*_0 : \ruleout{\varnothing}{\tau_0} \transto e^*_0$ hold, which brings to the arrow type in the thesis by rule \rulename{Abs}.
\qed
\end{lemma}

\begin{lemma}[Correctness of Inject-All Translation]
\label{lem:inject-all-translation}
Given a type derivation $\Gamma \vdash e_0 : \ruleout{\{ \many{x} : \many{\tau} \}}{\tau_0} \transto e_0^*$
and the derivation of the full injection
$\Gamma \vdash \injectop ~ e_0 : \ruleout{\varnothing}{\{ \many{x} : \many{\tau} ~|~ \rowtail \} \rightarrow \tau_0} \transto e^*$, 
then a derivation $\Gamma \vdash e^* : \ruleout{\varnothing}{\{ \many{x} : \many{\tau} ~|~ \rowtail \} \rightarrow \tau_0} \transto e^*$ exists.

\proof (Sketched)
Rule \rulename{Inject-All} treats full injection as syntactic sugar for restricted injection of the whole constraint set.
The thesis comes directly from Lemma \ref{lem:restricted-inject-translation}.
\qed
\end{lemma}

\begin{lemma}[Correctness of Eject-All Translation]
\label{lem:eject-all-translation}
Given the derivation
$\Gamma \vdash e_0 : \ruleout{\pi}{\{ \many{x} : \many{\tau} ~|~ \rowtail \} \rightarrow \tau_0} \transto e_0^*$
and the derivation of the full ejection
$\Gamma \vdash \ejectop ~ e_0 : \ruleout{\{ \many{x} : \many{\tau} \} \cup \pi}{\tau_0} \transto e^*$, 
then a derivation $\Gamma \cup \pi \vdash e^* : \ruleout{\varnothing}{\tau_0} \transto e^*$ exists.

\proof (Sketched)
From rule \rulename{Eject-All} comes the translation $e^* = e^*_0 ~ \{~ x_1 = x_1; ~ ..; ~ x_n = x_n ~\}$.
Also, the translated expression $e^*_0$ has the same type of $e_0$, i.e. $\{ \many{x} : \many{\tau} ~|~ \rowtail \} \rightarrow \tau_0$, except for free overloaded identifiers collected in $\pi$.
This means that the derivation $\Gamma \cup \pi \vdash e^*_0 : \ruleout{\varnothing}{\{ \many{x} : \many{\tau} ~|~ \rowtail \} \rightarrow \tau_0} \transto e^*_0$ holds.
By rules \rulename{Record} and \rulename{App} the record application erases the arrow type and $\tau_0$ remains.
\qed
\end{lemma}

\begin{lemma}[Correctness of Eject-R Translation]
\label{lem:restricted-eject-translation}
Given the derivation
$\Gamma \vdash e_0 : \ruleout{\pi}{\{ \many{x} : \many{\tau_x}; ~ \many{y} : \many{\tau_y} ~|~ \rowtail \} \rightarrow \tau_0} \transto e_0^*$
and the derivation of the restricted ejection
$\Gamma \vdash \ejectop ~ e_0 : \ruleout{\{ ~ \many{x} : \many{\tau_x} \} \cup \pi}{\{ \many{y} : \many{\tau_y} ~|~ \rowtail \} \rightarrow \tau_0} \transto e^*$, 
then a derivation $\Gamma \cup \{ \many{x} : \many{\tau_x} \} \cup \pi \vdash e^* : \ruleout{\varnothing}{\{ \many{y} : \many{\tau_y} ~|~ \rowtail \} \rightarrow \tau_0} \transto e^*$ exists.

\proof (Sketched)
This comes directly from rule \rulename{Eject-R} which is a syntactic sugar of a complete ejection and a restricted injection of remaining identifiers $\many{y}$.
From Lemma \ref{lem:eject-all-translation} comes that $\Gamma \cup \{ \many{x} : \many{\tau_x} \} \cup \{ \many{y} : \many{\tau_y} \} \cup \pi \vdash e^* : \ruleout{\varnothing}{\tau_0} \transto e^*$ due to full ejection.
Then from Lemma \ref{lem:restricted-inject-translation} comes that $\Gamma \cup \{ \many{x} : \many{\tau_x} \} \cup \pi \vdash e^* : \ruleout{\varnothing}{\{ \many{y} : \many{\tau_y} ~|~ \rowtail \} \rightarrow \tau_0} \transto e^*$ due to restricted re-injection of $\many{y}$.
\qed
\end{lemma}

\section{Conclusions}
\label{sec:conclusions}

We presented a language design for the easy integration of two orthogonal programming styles: manual dictionary passing and compiler-driven resolution of overloading constraints.
Rather than operating with heavyweight records and type classes, from which name clashing and other inconveniences would arise, our proposal is based on extensible records and fine-grained overloading constraints. 
On top of those, we introduce two special language operators, \injectop and \ejectop, converting constraints into record arguments and vice versa, allowing the programmer to switch between the two styles at no cost in terms of code rewriting.
Additionally, injection and ejection can also be restricted to a set of identifiers, allowing manipulation of parts of records or parts of the constraint set.

The design space explored here is not entirely novel, though the full potential of the reversible mechanism provided by combining injection and ejection remains to be discovered from a programmatic perspective.
The usefulness of code reuse across different programming styles is something that remains to be inspected.

As far as the performance of the proposed system is concerned, both injection and ejection revolve around dictionary passing, either manual or automatic, which ultimately reduces to the same mechanism Haskell and other languages supporting parametric ah-hoc polymorphism implement. The same performance is therefore expected. The cost of a constrained function compared to the cost of a non-constrained function is negligible, as it depends on the number of extra lambda parameters introduced by the constraints and the applications introduced for passing dictionaries once the constraints are solved. In other words, it is a cost in terms of argument passing that would take place anyway in the presence of any overloading system, with or without injection and ejection.

An experimental implementation of \injectop and \ejectop exists and is one of the central features of the functional programming language Lw (short form for Lightweight), whose highlight is a full lightweight approach: every language feature does not require annotations of declarations, unless explicitly desired by the programmer, while keeping a strong type discipline, type inference and other advanced features\footnote{Lw can be found on GitHub at the following URL: \url{https://github.com/alvisespano/Lw}. It is actively under development.}.

Injection and ejection require additional study and a more rigorous formalization into a theoretical framework to fully understand their implications.
A full proof of soundness is necessary as well as the preservation of semantics in the presence of injection/ejection, as well as a proof of coherence for the underlying overloading subsystem upon within our system relies.

As potential future work, we anticipate two enhancements that are worth investigating.

\subsection{As first-class operators}

Promoting \injectop and \ejectop to first-class citizens of the expression syntax would render their types polymorphic:
$$
\begin{array}{rcl}
\src{inject} & : & \forall \kindannot[\rowkind]{\alpha} ~ \kindannot[\rowkind]{\beta} ~ \kindannot{\gamma} . ~ (\beta \Rightarrow \gamma) \rightarrow (\{~ \beta ~|~ \alpha ~\} \rightarrow \gamma)
\\
\src{eject} & : & \forall \kindannot[\rowkind]{\alpha} ~ \kindannot[\rowkind]{\beta} ~ \kindannot{\gamma} . ~ (\{~ \beta ~|~ \alpha ~\} \rightarrow \gamma) \rightarrow (\beta \Rightarrow \gamma)
\end{array}
$$

That would allow functional composition and additional expressivity.
For instance, $\injectop \cdot \ejectop$ would lead to the identity function.
Also, \injectop and \ejectop could be passed as arguments to higher-order functions.
For example, given a list of functions picking a record argument, say \src{[f1; f2; f3]}, the expression \src{map eject [f1; f2; f3]} is capable of converting into constrained functions a whole library based on dictionary passing.

One major implication is that constrained types become first-class citizens of the type syntax, which complicates the type system substantially. First-class type classes \cite{Sozeau:2008:FTC:1459784.1459810} are not enough with our fine-grained constraint system.
Also, row types should be used to represent extensible records as well as constraint sets, which is something uncovered in the literature. 
Proposals for unifying type constraints and record types, such as \cite{gaster1998records}, are based on qualified types à la Haskell, which consist of heavyweight type classes and constraint names, rather than fine-grained overloaded symbols.

Another major challenge would be combining all this with first-class polymorphism systems such as MLF \cite{LeBotlan:2014:MRM:2641638.2641653} and HML \cite{Leijen:2009:FTR:1480881.1480891}, allowing System-F types within constraints as well as constraints within System-F types.

\subsection{Supporting extensible records with overloaded labels}

One limitation of the system proposed in this paper, specifically of injection, emerges when the same overloaded symbol appears multiple times within a constraint set.
When injected, that would be converted into a record with multiple labels with the same name, which requires extensible records with overloadable labels.
Take a simple pretty printer for a record type.

\begin{lwcode}
overload pretty : 'a -> string

let pretty_person p = (pretty p.name) ^ ", " ^ (pretty p.age)

let pp = inject pretty_person
\end{lwcode}

Here the inferred type is $\src{pretty\_person} : \{~ \src{pretty} : \alpha \rightarrow \src{string}; ~ \src{pretty} : \beta \rightarrow \src{string} ~\} \Rightarrow \{~ \src{name} : \alpha; ~ \src{age} : \beta ~|~ \kindannot[\rowkind]{\gamma} ~\} \rightarrow \src{string}$ because there are two distinct occurrences of the overloaded symbol \src{pretty}.
This is standard behaviour in open-world overloading systems: each different occurrence is internally represented with a suffix appended to the base identifier name in order to distinguish between them.
We discussed this mechanism in Section \ref{sec:type-system}.

Complications arise when injection comes into play.
The type of $\src{pp} : \{~ \src{pretty} : \alpha \rightarrow \src{string}; ~ \src{pretty} : \beta \rightarrow \src{string} ~\} \rightarrow \{~ \src{name} : \alpha; ~ \src{age} : \beta ~|~ \kindannot[\rowkind]{\gamma} ~\} \rightarrow \src{string}$ reveals that the two occurrences of \src{pretty} have now become two record fields with the same overloaded name, which is something not supported by the extensible record system we adopted \cite{extensible-records-with-scoped-labels} - or by any other extensible record system, for that matter.
Scoped access to labels allows for multiple fields with the same name to co-exist, though that does not mean they are overloaded.

GHC supports overloaded record labels by means of a sophisticated use of type classes, though that applies only to heavyweight record types.
Implementing overloading of labels for an extensible record system is something uncovered in the literature, and the reason is that it introduces a borderline design.
Misuses may emerge easily:
\begin{lwcode}
let f r = if r.a then r.a + 1 else r.a r.b
\end{lwcode}

Record parameter $\src{r} : \{~ \src{a} : \src{bool}; ~ \src{a} : \src{int}; ~ \src{a} : \alpha \rightarrow \src{int}; ~ \src{b} : \alpha ~\}$ contains a field $a$ appearing multiple times with different types, which is pretty quirky.
Such an improvement would however render injection fully functional in conjunction with multiple occurrences of the same constraint name, hence ejection fully reversible in such extreme cases.





\appendix


\bibliographystyle{elsarticle-num-names} 
\bibliography{local}

\begin{thebibliography}{39}
\expandafter\ifx\csname natexlab\endcsname\relax\def\natexlab#1{#1}\fi
\providecommand{\url}[1]{\texttt{#1}}
\providecommand{\href}[2]{#2}
\providecommand{\path}[1]{#1}
\providecommand{\DOIprefix}{doi:}
\providecommand{\ArXivprefix}{arXiv:}
\providecommand{\URLprefix}{URL: }
\providecommand{\Pubmedprefix}{pmid:}
\providecommand{\doi}[1]{\href{http://dx.doi.org/#1}{\path{#1}}}
\providecommand{\Pubmed}[1]{\href{pmid:#1}{\path{#1}}}
\providecommand{\bibinfo}[2]{#2}
\ifx\xfnm\relax \def\xfnm[#1]{\unskip,\space#1}\fi
\bibitem[{Cardelli and Wegner(1985)}]{cardelli1985understanding}
\bibinfo{author}{L.~Cardelli}, \bibinfo{author}{P.~Wegner},
\newblock \bibinfo{title}{On understanding types, data abstraction, and polymorphism},
\newblock \bibinfo{journal}{ACM Computing Surveys (CSUR)} \bibinfo{volume}{vol. 17} (\bibinfo{year}{1985}) \bibinfo{pages}{pp. 471--523}.
\bibitem[{Cardelli and Mitchell(1990)}]{Cardelli1990}
\bibinfo{author}{L.~Cardelli}, \bibinfo{author}{J.~C. Mitchell}, \bibinfo{title}{Operations on records}, \bibinfo{publisher}{Springer New York}, \bibinfo{address}{New York, NY}, \bibinfo{year}{1990}, pp. \bibinfo{pages}{22--52}.
\bibitem[{Cardelli(1992)}]{cardelli1992extensible}
\bibinfo{author}{L.~Cardelli}, \bibinfo{title}{Extensible records in a pure calculus of subtyping}, \bibinfo{publisher}{Digital. Systems Research Center}, \bibinfo{year}{1992}.
\bibitem[{Jategaonkar and Mitchell(1993)}]{jategaonkar1993type}
\bibinfo{author}{L.~A. Jategaonkar}, \bibinfo{author}{J.~C. Mitchell},
\newblock \bibinfo{title}{Type inference with extended pattern matching and subtypes},
\newblock \bibinfo{journal}{Fundamenta Informaticae} \bibinfo{volume}{vol. 19} (\bibinfo{year}{1993}) \bibinfo{pages}{pp. 127--165}.
\bibitem[{Wand(1987)}]{Wand1987CompleteTI}
\bibinfo{author}{M.~Wand},
\newblock \bibinfo{title}{Complete type inference for simple objects},
\newblock in: \bibinfo{booktitle}{LICS}, \bibinfo{year}{1987}.
\bibitem[{Wand(1991)}]{wand1991type}
\bibinfo{author}{M.~Wand},
\newblock \bibinfo{title}{Type inference for record concatenation and multiple inheritance},
\newblock \bibinfo{journal}{Information and Computation} \bibinfo{volume}{vol. 93} (\bibinfo{year}{1991}) \bibinfo{pages}{pp. 1--15}.
\bibitem[{R{\'e}my(1989)}]{Remy:1989:TCR:75277.75284}
\bibinfo{author}{D.~R{\'e}my},
\newblock \bibinfo{title}{Type checking records and variants in a natural extension of ml},
\newblock in: \bibinfo{booktitle}{Proceedings of the 16th ACM SIGPLAN-SIGACT Symposium on Principles of Programming Languages}, POPL '89, \bibinfo{publisher}{ACM New York, NY, USA}, \bibinfo{year}{1989}, pp. \bibinfo{pages}{77--88}.
\bibitem[{Jones and Jones(1999)}]{jones1999lightweight}
\bibinfo{author}{M.~P. Jones}, \bibinfo{author}{S.~P. Jones},
\newblock \bibinfo{title}{Lightweight extensible records for haskell},
\newblock in: \bibinfo{booktitle}{Haskell Workshop}, \bibinfo{organization}{Citeseer}, \bibinfo{year}{1999}.
\bibitem[{Gaster and Jones(1996)}]{gaster1996polymorphic}
\bibinfo{author}{B.~R. Gaster}, \bibinfo{author}{M.~P. Jones}, \bibinfo{title}{A polymorphic type system for extensible records and variants}, \bibinfo{type}{Technical Report}, Technical Report NOTTCS-TR-96-3, Department of Computer Science, University of Nottingham, \bibinfo{year}{1996}.
\bibitem[{Gaster(1998)}]{gaster1998records}
\bibinfo{author}{B.~R. Gaster}, \bibinfo{title}{Records, variants and qualified types}, Ph.D. thesis, Citeseer, \bibinfo{year}{1998}.
\bibitem[{Kaes(1988)}]{kaes1988parametric}
\bibinfo{author}{S.~Kaes},
\newblock \bibinfo{title}{Parametric overloading in polymorphic programming languages},
\newblock in: \bibinfo{booktitle}{ESOP'88: 2nd European Symposium on Programming Nancy, France, March 21--24, 1988 Proceedings 2}, \bibinfo{organization}{Springer}, \bibinfo{year}{1988}, pp. \bibinfo{pages}{131--144}.
\bibitem[{Wadler and Blott(1989)}]{Wadler:1989:MAP:75277.75283}
\bibinfo{author}{P.~Wadler}, \bibinfo{author}{S.~Blott},
\newblock \bibinfo{title}{How to make ad-hoc polymorphism less ad hoc},
\newblock in: \bibinfo{booktitle}{Proceedings of the 16th ACM SIGPLAN-SIGACT Symposium on Principles of Programming Languages}, POPL '89, \bibinfo{publisher}{ACM New York, NY, USA}, \bibinfo{year}{1989}, pp. \bibinfo{pages}{60--76}.
\bibitem[{Jones(1992)}]{jones1992theory}
\bibinfo{author}{M.~P. Jones},
\newblock \bibinfo{title}{A theory of qualified types},
\newblock in: \bibinfo{booktitle}{European symposium on programming}, \bibinfo{organization}{Springer}, \bibinfo{year}{1992}, pp. \bibinfo{pages}{287--306}.
\bibitem[{Jones(1995)}]{Jones:1995:SIQ:224164.224198}
\bibinfo{author}{M.~P. Jones},
\newblock \bibinfo{title}{Simplifying and improving qualified types},
\newblock in: \bibinfo{booktitle}{Proceedings of the Seventh International Conference on Functional Programming Languages and Computer Architecture}, FPCA '95, \bibinfo{publisher}{ACM New York, NY, USA}, \bibinfo{year}{1995}, pp. \bibinfo{pages}{160--169}.
\bibitem[{Hall et~al.(1996)Hall, Hammond, Peyton~Jones, and Wadler}]{hall1996type}
\bibinfo{author}{C.~V. Hall}, \bibinfo{author}{K.~Hammond}, \bibinfo{author}{S.~L. Peyton~Jones}, \bibinfo{author}{P.~L. Wadler},
\newblock \bibinfo{title}{Type classes in haskell},
\newblock \bibinfo{journal}{ACM Transactions on Programming Languages and Systems (TOPLAS)} \bibinfo{volume}{vol. 18} (\bibinfo{year}{1996}) \bibinfo{pages}{pp. 109--138}.
\bibitem[{Jones et~al.(1997)Jones, Jones, and Meijer}]{Jones97typeclasses}
\bibinfo{author}{S.~P. Jones}, \bibinfo{author}{M.~Jones}, \bibinfo{author}{E.~Meijer},
\newblock \bibinfo{title}{Type classes: an exploration of the design space},
\newblock in: \bibinfo{booktitle}{Haskell workshop}, \bibinfo{year}{1997}, pp. \bibinfo{pages}{1--16}.
\bibitem[{Odersky et~al.(2004)Odersky, Altherr, Cremet, Emir, Maneth, Micheloud, Mihaylov, Schinz, Stenman, and Zenger}]{odersky2004overview}
\bibinfo{author}{M.~Odersky}, \bibinfo{author}{P.~Altherr}, \bibinfo{author}{V.~Cremet}, \bibinfo{author}{B.~Emir}, \bibinfo{author}{S.~Maneth}, \bibinfo{author}{S.~Micheloud}, \bibinfo{author}{N.~Mihaylov}, \bibinfo{author}{M.~Schinz}, \bibinfo{author}{E.~Stenman}, \bibinfo{author}{M.~Zenger},
\newblock \bibinfo{title}{An overview of the scala programming language}  (\bibinfo{year}{2004}).
\bibitem[{Odersky and Rompf(2014)}]{Odersky2014}
\bibinfo{author}{M.~Odersky}, \bibinfo{author}{T.~Rompf},
\newblock \bibinfo{title}{Unifying functional and object-oriented programming with scala},
\newblock \bibinfo{journal}{Communications of the ACM} \bibinfo{volume}{vol. 57} (\bibinfo{year}{2014}) \bibinfo{pages}{pp. 76--86}.
\bibitem[{Camar\~{a}o and Figueiredo(1999)}]{Camarao:1999:TIO:646189.683411}
\bibinfo{author}{C.~Camar\~{a}o}, \bibinfo{author}{L.~Figueiredo},
\newblock \bibinfo{title}{Type inference for overloading without restrictions, declarations or annotations},
\newblock in: \bibinfo{booktitle}{Proceedings of the 4th Fuji International Symposium on Functional and Logic Programming}, FLOPS '99, \bibinfo{publisher}{Springer-Verlag}, \bibinfo{address}{London, UK}, \bibinfo{year}{1999}, pp. \bibinfo{pages}{37--52}.
\bibitem[{Camar\~{a}o et~al.(2004)Camar\~{a}o, Figueiredo, and Vasconcellos}]{Camarao:2004:CSO:1013963.1013974}
\bibinfo{author}{C.~Camar\~{a}o}, \bibinfo{author}{L.~Figueiredo}, \bibinfo{author}{C.~Vasconcellos},
\newblock \bibinfo{title}{Constraint-set satisfiability for overloading},
\newblock in: \bibinfo{booktitle}{Proceedings of the 6th ACM SIGPLAN International Conference on Principles and Practice of Declarative Programming}, PPDP '04, \bibinfo{publisher}{ACM New York, NY, USA}, \bibinfo{year}{2004}, pp. \bibinfo{pages}{67--77}.
\bibitem[{Odersky et~al.(1995)Odersky, Wadler, and Wehr}]{Odersky:1995:SLO:224164.224195}
\bibinfo{author}{M.~Odersky}, \bibinfo{author}{P.~Wadler}, \bibinfo{author}{M.~Wehr},
\newblock \bibinfo{title}{A second look at overloading},
\newblock in: \bibinfo{booktitle}{Proceedings of the Seventh International Conference on Functional Programming Languages and Computer Architecture}, FPCA '95, \bibinfo{publisher}{ACM New York, NY, USA}, \bibinfo{year}{1995}, pp. \bibinfo{pages}{135--146}.
\bibitem[{Bruce et~al.(1999)Bruce, Cardelli, and Pierce}]{bruce1999comparing}
\bibinfo{author}{K.~B. Bruce}, \bibinfo{author}{L.~Cardelli}, \bibinfo{author}{B.~C. Pierce},
\newblock \bibinfo{title}{Comparing object encodings},
\newblock \bibinfo{journal}{Information and Computation} \bibinfo{volume}{vol. 155} (\bibinfo{year}{1999}) \bibinfo{pages}{pp. 108--133}.
\bibitem[{Devriese and Piessens(2011)}]{Devriese:2011:BST:2034773.2034796}
\bibinfo{author}{D.~Devriese}, \bibinfo{author}{F.~Piessens},
\newblock \bibinfo{title}{On the bright side of type classes: Instance arguments in agda},
\newblock in: \bibinfo{booktitle}{Proceedings of the 16th ACM SIGPLAN International Conference on Functional Programming}, ICFP '11, \bibinfo{publisher}{ACM New York, NY, USA}, \bibinfo{year}{2011}, pp. \bibinfo{pages}{143--155}.
\bibitem[{Winant and Devriese(2018)}]{winant2018coherent}
\bibinfo{author}{T.~Winant}, \bibinfo{author}{D.~Devriese},
\newblock \bibinfo{title}{Coherent explicit dictionary application for haskell},
\newblock \bibinfo{journal}{ACM SIGPLAN Notices} \bibinfo{volume}{vol. 53} (\bibinfo{year}{2018}) \bibinfo{pages}{pp. 81--93}.
\bibitem[{Leijen(2005)}]{extensible-records-with-scoped-labels}
\bibinfo{author}{D.~Leijen},
\newblock \bibinfo{title}{Extensible records with scoped labels},
\newblock in: \bibinfo{booktitle}{Proceedings of the 2005 Symposium on Trends in Functional Programming (TFP'05), Tallin, Estonia}, \bibinfo{year}{2005}.
\bibitem[{Lewis et~al.(2000)Lewis, Launchbury, Meijer, and Shields}]{Lewis:2000:IPD:325694.325708}
\bibinfo{author}{J.~R. Lewis}, \bibinfo{author}{J.~Launchbury}, \bibinfo{author}{E.~Meijer}, \bibinfo{author}{M.~B. Shields},
\newblock \bibinfo{title}{Implicit parameters: Dynamic scoping with static types},
\newblock in: \bibinfo{booktitle}{Proceedings of the 27th ACM SIGPLAN-SIGACT Symposium on Principles of Programming Languages}, POPL '00, \bibinfo{publisher}{ACM New York, NY, USA}, \bibinfo{year}{2000}, pp. \bibinfo{pages}{108--118}.
\bibitem[{Sozeau and Oury(2008)}]{Sozeau:2008:FTC:1459784.1459810}
\bibinfo{author}{M.~Sozeau}, \bibinfo{author}{N.~Oury},
\newblock \bibinfo{title}{First-class type classes},
\newblock in: \bibinfo{booktitle}{Proceedings of the 21st International Conference on Theorem Proving in Higher Order Logics}, TPHOLs '08, \bibinfo{publisher}{Springer-Verlag}, \bibinfo{address}{Berlin, Heidelberg}, \bibinfo{year}{2008}, pp. \bibinfo{pages}{278--293}.
\bibitem[{Dijkstra et~al.(2005)Dijkstra, Swierstra et~al.}]{dijkstra2005making}
\bibinfo{author}{A.~Dijkstra}, \bibinfo{author}{S.~D. Swierstra}, et~al., \bibinfo{title}{Making implicit parameters explicit}, \bibinfo{type}{Technical Report}, Technical report, Utrecht University, \bibinfo{year}{2005}.
\bibitem[{Kahl and Scheffczyk(2001)}]{kahl2001named}
\bibinfo{author}{W.~Kahl}, \bibinfo{author}{J.~Scheffczyk},
\newblock \bibinfo{title}{Named instances for haskell type classes},
\newblock in: \bibinfo{booktitle}{Proceedings of the 2001 Haskell Workshop, number UU-CS-2001-23 in Tech. Rep}, \bibinfo{year}{2001}, pp. \bibinfo{pages}{71--99}.
\bibitem[{Jones(1993)}]{jones1993coherence}
\bibinfo{author}{M.~P. Jones}, \bibinfo{title}{Coherence for qualified types}, \bibinfo{type}{Technical Report}, Technical Report YALEU/DCS/RR-989, Yale University, \bibinfo{year}{1993}.
\bibitem[{Bottu et~al.(2019)Bottu, Xie, Marntirosian, and Schrijvers}]{bottu2019coherence}
\bibinfo{author}{G.-J. Bottu}, \bibinfo{author}{N.~Xie}, \bibinfo{author}{K.~Marntirosian}, \bibinfo{author}{T.~Schrijvers},
\newblock \bibinfo{title}{Coherence of type class resolution},
\newblock \bibinfo{journal}{Proceedings of the ACM on Programming Languages} \bibinfo{volume}{vol. 3} (\bibinfo{year}{2019}) \bibinfo{pages}{pp. 1--28}.
\bibitem[{Schrijvers et~al.(2019)Schrijvers, Oliveira, Wadler, and Marntirosian}]{schrijvers2019cochis}
\bibinfo{author}{T.~Schrijvers}, \bibinfo{author}{B.~C. Oliveira}, \bibinfo{author}{P.~Wadler}, \bibinfo{author}{K.~Marntirosian},
\newblock \bibinfo{title}{Cochis: Stable and coherent implicits},
\newblock \bibinfo{journal}{Journal of Functional Programming} \bibinfo{volume}{vol. 29} (\bibinfo{year}{2019}).
\bibitem[{Stuckey and Sulzmann(2005)}]{stuckey2005theory}
\bibinfo{author}{P.~J. Stuckey}, \bibinfo{author}{M.~Sulzmann},
\newblock \bibinfo{title}{A theory of overloading},
\newblock \bibinfo{journal}{ACM Transactions on Programming Languages and Systems (TOPLAS)} \bibinfo{volume}{vol. 27} (\bibinfo{year}{2005}) \bibinfo{pages}{pp. 1216--1269}.
\bibitem[{Duggan and Ophel(2002)}]{duggan2002open}
\bibinfo{author}{D.~Duggan}, \bibinfo{author}{J.~Ophel},
\newblock \bibinfo{title}{Open and closed scopes for constrained genericity},
\newblock \bibinfo{journal}{Theoretical Computer Science} \bibinfo{volume}{vol. 275} (\bibinfo{year}{2002}) \bibinfo{pages}{pp. 215--258}.
\bibitem[{Oliveira et~al.(2010)Oliveira, Moors, and Odersky}]{Oliveira:2010:TCO:1932682.1869489}
\bibinfo{author}{B.~C. Oliveira}, \bibinfo{author}{A.~Moors}, \bibinfo{author}{M.~Odersky},
\newblock \bibinfo{title}{Type classes as objects and implicits},
\newblock \bibinfo{journal}{ACM SIGPLAN Notices} \bibinfo{volume}{vol. 45} (\bibinfo{year}{2010}) \bibinfo{pages}{pp. 341--360}.
\bibitem[{Oliveira et~al.(2012)Oliveira, Schrijvers, Choi, Lee, and Yi}]{oliveira2012implicit}
\bibinfo{author}{B.~C. Oliveira}, \bibinfo{author}{T.~Schrijvers}, \bibinfo{author}{W.~Choi}, \bibinfo{author}{W.~Lee}, \bibinfo{author}{K.~Yi},
\newblock \bibinfo{title}{The implicit calculus: a new foundation for generic programming},
\newblock \bibinfo{journal}{ACM SIGPLAN Notices} \bibinfo{volume}{vol. 47} (\bibinfo{year}{2012}) \bibinfo{pages}{pp. 35--44}.
\bibitem[{Jones(2000)}]{jones2000type}
\bibinfo{author}{M.~P. Jones},
\newblock \bibinfo{title}{Type classes with functional dependencies},
\newblock in: \bibinfo{booktitle}{European Symposium on Programming}, \bibinfo{organization}{Springer}, \bibinfo{year}{2000}, pp. \bibinfo{pages}{230--244}.
\bibitem[{Leijen(2009)}]{Leijen:2009:FTR:1480881.1480891}
\bibinfo{author}{D.~Leijen},
\newblock \bibinfo{title}{Flexible types: Robust type inference for first-class polymorphism},
\newblock in: \bibinfo{booktitle}{Proceedings of the 36th Annual ACM SIGPLAN-SIGACT Symposium on Principles of Programming Languages}, POPL '09, \bibinfo{publisher}{ACM New York, NY, USA}, \bibinfo{year}{2009}, pp. \bibinfo{pages}{66--77}.
\bibitem[{Le~Botlan and R{\'e}my(2014)}]{LeBotlan:2014:MRM:2641638.2641653}
\bibinfo{author}{D.~Le~Botlan}, \bibinfo{author}{D.~R{\'e}my},
\newblock \bibinfo{title}{Mlf: raising ml to the power of system f},
\newblock \bibinfo{journal}{ACM SIGPLAN Notices} \bibinfo{volume}{vol. 49} (\bibinfo{year}{2014}) \bibinfo{pages}{pp. 52--63}.

\end{thebibliography}





\end{document}